\documentclass[]{aa}  

\usepackage{graphicx}
\usepackage{txfonts}
\usepackage{subfig}
\usepackage{textgreek}
\usepackage{multirow}
\usepackage{mathtools}
\usepackage{stfloats}
\usepackage{color}
\usepackage{rotating}
\usepackage{enumitem}

\bibpunct{(}{)}{;}{a}{}{,}

\newcommand{\FIG}[1]{}

\def\mso{\,{\rm M}_\odot}

\def\rso{\,{\rm R}_\odot}

\def\kms{\, {\rm km}\, {\rm s}^{-1}}

\def\msoy{\, \mso~{\rm yr}^{-1}}

\begin{document}

	
	\title{{\sc Atomium}: A high-resolution view on the highly asymmetric wind of the AGB star $\pi^1$Gruis}
	
	\subtitle{I. First detection of a new companion and its effect on the inner wind}
	
	\author{Ward Homan
		\inst{1}
		\and
		Miguel Montarg\`{e}s
		\inst{1}
		\and
		Bannawit Pimpanuwat
		\inst{2}
		\and
		Anita M. S. Richards
		\inst{2}
		\and
		Sofia H. J. Wallstr\"om
		\inst{1}
		\and
		Pierre Kervella
		\inst{4}
		\and          
		Leen Decin
		\inst{1,3}
		\and   
		Albert Zijlstra
		\inst{2}
		\and
		Taissa Danilovich
		\inst{1}
		\and
		Alex de Koter
		\inst{5,1}
		\and
		Karl Menten
		\inst{6}
		\and
		Raghvendra Sahai
		\inst{7}
		\and
		John Plane
		\inst{3}
		\and
		Kelvin Lee
		\inst{9}
		\and
		Rens Waters
		\inst{5}
		\and
		Alain Baudry
		\inst{10}
		\and
		Ka Tat Wong
		\inst{11}
		\and
		Tom J. Millar
		\inst{12}
		\and
		Marie Van de Sande
		\inst{1}
		\and
		Eric Lagadec
		\inst{13}
		\and
		David Gobrecht
		\inst{1}
		\and
		Jeremy Yates
		\inst{14}
		\and
		Daniel Price
		\inst{15}
		\and
		Emily Cannon
		\inst{1}
		\and
		Jan Bolte
		\inst{1}
		\and
		Frederik De Ceuster
		\inst{1,14}
		\and
		Fabrice Herpin
		\inst{10}
		\and
		Joe Nuth
		\inst{16}
		\and
	    Jan Philip Sindel
		\inst{1}
		\and
		Dylan Kee
		\inst{1}
		\and
		Malcolm D. Grey
		\inst{2,20}
		\and
		Sandra Etoka
		\inst{2}
		\and
		Manali Jeste
		\inst{6}
		\and
		Carl A. Gottlieb
		\inst{17}
		\and
		Elaine Gottlieb
		\inst{21}
		\and
		Iain McDonald
		\inst{2}
		\and
		Ileyk El Mellah
		\inst{18}
		\and
		Holger S. P. Müller
		\inst{19}
	}
	
	\offprints{W. Homan}          
	
	\institute{
	    $^{\rm 1}\ $Institute of Astronomy, KU Leuven, Celestijnenlaan 200D B2401, 3001 Leuven, BE \\
		$^{\rm 2}\ $JBCA, Department Physics and Astronomy, University of Manchester, Manchester M13 9PL, UK \\
		$^{\rm 3}\ $School of Chemistry, University of Leeds, Leeds LS2 9JT, UK \\   
		$^{\rm 4}\ $LESIA (CNRS UMR 8109), Observatoire de Paris, PSL, CNRS, UPMC, Univ. Paris-Diderot, FR \\   
		$^{\rm 5}\ $Astronomical Institute Anton Pannekoek, University of Amsterdam, Science Park 904, PO Box 94249, 1090 GE, Amsterdam, NL \\   
		$^{\rm 6}\ $Max-Planck-Institut für Radioastronomie, auf dem Hügel 69, 53121 Bonn, GE \\
		$^{\rm 7}\ $Jet Propulsion Laboratory, California Institute of Technology, 4800 Oak Grove Drive, Pasadena, CA 91109, USA \\
		$^{\rm 9}\ $Radio and Geoastronomy Division, Harvard-Smithsonian Center for Astrophysics, Cambridge, MA, USA \\
		$^{\rm 10}\ $Laboratoire d'astrophysique de Bordeaux, Université de Bordeaux, CNRS, B18N, Allée Geoffroy Saint-Hilaire, 33615 Pessac, FR. \\
		$^{\rm 11}\ $Institut de Radioastronomie Millimétrique, 300 rue de la Piscine, 38406 Saint Martin d’Hères, FR \\
		$^{\rm 12}\ $Astrophysics Research Centre, School of Mathematics and Physics, Queen’s University Belfast, University Road, Belfast BT7 1NN, UK. \\
		$^{\rm 13}\ $Laboratoire Lagrange, Université Côte d’Azur, Observatoire de la Côte d’Azur, CNRS, Boulevard de l’Observatoire, CS 34229, F-06304 Nice Cedex 4, FR. \\
		$^{\rm 14}\ $Department of Physics and Astronomy, University College London, Gower Street London, WC1E 6BT, UK \\
		$^{\rm 15}\ $School of Physics \& Astronomy, Monash University, Clayton, Vic 3800, auS \\
		$^{\rm 16}\ $Solar System Exploration Division, Code 690 NASA’s Goddard Space Flight Center, Greenbelt MD 20771 USA \\
		$^{\rm 17}\ $Harvard-Smithsonian Center for Astrophysics, 60 Garden Street Cambridge, MA 02138, USA \\
		$^{\rm 18}\ $Center for Mathematical Plasma Astrophysics, Celestijnenlaan 200B, 3001 Leuven, BE \\
		$^{\rm 19}\ $I. Physikalisches Institut, Universität zu Köln, Zülpicher Str. 77, 50937 Köln, GE \\
		$^{\rm 20}\ $National Astronomical Research Institute of Thailand, 260 Moo 4, T. Donkaew, A. Maerim, Chiangmai 50180, TH.\\
		$^{\rm 21}\ $School of Engineering and Applied Sciences and Department of Earth and Planetary Sciences, Harvard University, USA.
	}             
	
	\date{Received <date> / Accepted <date>}
	
	\abstract  
	{The nebular circumstellar environments of cool evolved stars are known to harbour a rich morphological complexity of gaseous structures on different length scales. A large part of these density structures are thought to be brought about by the interaction of the stellar wind with a close companion. The S-type asymptotic giant branch star $\pi^1$Gruis, which has a known companion at $\sim$440~au and is thought to harbour a second, closer-by (<10~au) companion, was observed with the Atacama Large Millimeter/submillimeter Array as part of the {\sc {\sc Atomium}} Large programme. In this work, the brightest CO, SiO, and HCN molecular line transitions are analysed. The continuum map shows two maxima, separated by 0.04'' (6~au). The CO data unambiguously reveal that $\pi^1$Gru's circumstellar environment harbours an inclined, radially outflowing, equatorial density enhancement. It contains a spiral structure at an angle of $\sim$38+/-3$^\circ$ with the line-of-sight. The HCN emission in the inner wind reveals a clockwise spiral, with a dynamical crossing time of the spiral arms consistent with a companion at a distance of 0.04" from the asymptotic giant branch star, which is in agreement with the position of the secondary continuum peak. The inner wind dynamics imply a large acceleration region, consistent with a beta-law power of $\sim$6. The CO emission suggests that the spiral is approximately Archimedean within 5'', beyond which this trend breaks down as the succession of the spiral arms becomes less periodic. The SiO emission at scales smaller than 0.5'' exhibits signatures of gas in rotation, which is found to fit the expected behaviour of gas in the wind-companion interaction zone. An investigation of SiO maser emission reveals what could be a stream of gas accelerating from the surface of the AGB star to the companion. Using these dynamics, we have tentatively derived an upper limit on the companion mass to be $\sim$1.1~$\mso$.}
	
	
	\maketitle
	
	\section{Introduction} \label{intro}
	
	Before turning into extended and intricately shaped planetary nebulae (PNe), low and intermediate mass stars evolve up the asymptotic giant branch (AGB) in the Hertzsprung--Russell diagram. The AGB phase is characterised by a strong mass-loss event which spans a period of a few $10^5$ to $10^6$ years and completely strips the giant star of its convective atmosphere \citep{Habing2004}. This stellar wind is thought to be driven by stellar surface pulsations \citep{Bowen1988,Hofner2018,McDonald2019}, which lift gas to regions suitable for the formation of solid-state particles. The opacity of the dust species efficiently couples the dust to the incident stellar radiation, transferring a fraction of the radiation momentum to the dust particles and accelerating them radially outward \citep{Winters2000,Hofner2003,Woitke2006}. Drag forces applied by the dust to the surrounding gas results in the formation of a vast, nebulous circumstellar environment (CSE) that is rich in complex and interlinked thermal, dynamical, chemical, and radiative processes \citep{Hofner2018}.
	
	Though many of these CSEs seem to exhibit large-scale sphericity, they systematically display a wealth of smaller scale structural complexities, including arcs \citep[e.g.][]{Decin2012,Cox2012}, shells \citep[e.g.][]{Mauron2000}, clumps \citep[e.g.][]{Bowers1990,Ohnaka2016,Khouri2016}, spirals \citep[e.g.][]{Mauron2006,Mayer2011,Maercker2012,Kim2013}, voids \citep{Ramstedt2014}, tori \citep[e.g.][]{Skinner1998}, and rotating discs \citep{Kervella2016,Homan2018}. The nebulae surrounding semi-regular pulsator stars appear to be particularly susceptible to the production of axi-symmetric geometries. These are typically composed of a relatively flat equatorial density enhancement, which may or may not harbour a spiral-shaped density pattern, with a perpendicularly oriented bi-conical outflow \citep[e.g.][]{CastroCarrizo2010,Kervella2016,Doan2017,Homan2018,Homan2018b,Doan2020}
	
	Hydrodynamical calculations reveal that gravitational perturbations by a nearby companion can explain a variety of these morphologies \citep[e.g.][]{Soker1997,Huggins2007,Mastrodemos1999,Kim2011,Kim2012}. The companion frequency of the main sequence (MS) predecessors of AGB stars has been determined to exceed 50\%\ \citep{Raghavan2010,Duchene2013}, not including any planetary companions. Considering that recent studies show that on average every star in the Milky Way \citep[e.g.][]{Cassan2012} possesses one or more planets, this frequency can only be considered as a lower limit. Hence, the systematic occurrence of companion-induced perturbations within AGB stellar winds should not come as a surprise.
	
	The currently available high spatial resolution facilities present the unique possibility of studying the winds of these AGB stars in unprecedented detail. Imaging the CSE on milliarcsecond scales provides valuable insights into the local physics dominating the inner wind, which likely dictates the global shape and evolution of the CSE, and potentially even sheds some light on mechanisms that may (partly) explain further morphological evolution into post-AGB and planetary nebulae \citep{Decin2020}. Thus, imaging the complex structures in the inner regions of AGB CSEs at high angular resolution will aid our understanding of the missing morphological link between AGB stars and their PN progeny.
	
	This study focuses on $\pi^1$Gruis, an AGB star displaying semi-regular (SRb-type) variability \citep{Jorissen1993}, with a period of 195 days \citep{Pojmanski2005}. It is an intrinsic S-type star \citep{Jorissen1998}, which is characterised by the presence of ZrO bands and Tc lines in its photospheric emission, indicative of enhanced s-process element abundances. The stellar photospheres of S-type stars also show C/O abundance ratios close to unity \citep{Scalo1976,Smith1986}, which indicate that they are in the process of transitioning from an oxygen- to a carbon-dominated atmosphere. Its \emph{Hipparcos} parallax places it at a distance of approximately 163$\pm$20 parsec \citep{VanLeeuwen2007}. \citet{VanEck1998} estimated an effective temperature of $T_*\simeq$ 3100K and a luminosity of $\log (\mathrm{L}/\mathrm{L}_\odot)=3.86$. It has an estimated current mass of $M_*\simeq1.5\,\mso$ \citep{Siess2006,Siess2008}, a measured photospheric angular diameter of $\sim$20 milliarcseconds \citep{Paladini2018}, and a measured radial velocity relative to the local standard of rest (LSR) of $v_{*} = {\rm -12}\ \kms$ \citep{Danilovich2015}.
	
	The circumstellar environment of $\pi^1$Gru has been extensively documented in the literature. \citet{Sahai1992} discovered, using single-dish mapping of CO $J$=2-1 emission, a fast bipolar outflow (with speeds larger than 38~$\kms$) and an equatorial density enhancement (EDE) in this object. Resolved maps of $^{12}$CO $J$=2$-$1 emission were first made by \citet{Knapp1999} and later by \citet{Chiu2006}, at a resolution of $\sim$3''. They confirm the presence  of an expanding EDE at an inclination angle $i$ of 55$^\circ$ with respect to the line of sight.  We adopt the convention that 0$^\circ$ < $i$ < 90$^\circ$ implies that the southern side is closer to Earth. In 2017 $\pi^1$Gru was imaged using ALMA \citep{Doan2017} at a resolution of $\sim$5'', and modelled with the radiative transfer solver SHAPEMOL by \citep{Steffen2006}. Their analysis, which takes into consideration multiple rotational transitions and isotopologues of CO, constrains the velocity, density and temperature field within the EDE and the high-velocity polar outflows. From the mass in the EDE the authors deduce an upper limit on the mass-loss rate of $\sim7.7\times10^{-7}\msoy$. They visually constrain the inclination angle of the system to about 40$^\circ$. 
	
	$\pi^1$Gru is also a known binary system: \citet{Feast1953} and \citet{Ake1992} detected a G0V companion at a separation of 2.7'' ($\sim440$ astronomical units, henceforth au) from the central star. Considering the large orbital separation, \citet{Sahai1992} suspected that this companion could not be the driver behind the wind shaping, and that a closer by, to date undetected, companion would be required to explain the observed morphology. In later work, \citet{Mayer2014} modelled \emph{Herschel/PACS} and \emph{VLTI/AMBER} data to also conclude that the known, distant G0V companion cannot be responsible for the wind shaping. They deduced that a companion located at a distance of $<$10~au from the central star, with a tentative mass of $\sim$1~$\mso$, fitted the data well. Finally, \citet{Doan2020} analysed ALMA band 6 (211 -- 275~GHz) observations made with the 12m array in compact configuration, with a spatial resolution of $\sim$1.2''. They detect arcs in the nebular EDE which they interpret as being a counter-clockwise spiral. Additionally, they revealed that the fast outflow component has a bi-conical or hourglass-shaped morphology. Using 3D radiative transfer they construct a retrieval model for both the spiral and the hourglass, and also present a low resolution SPH model for a binary in an eccentric orbit with the aim of explaining certain second-order perturbations that appear in the data.
	
	As part of the {\sc Atomium} ALMA large programme (2018.1.00659.L. PI L. Decin. see \citep{Decin2020} and Gottlieb et al. \emph{in prep.} for an overview), we conducted spatially resolved cycle 6 ALMA band 6 observations of the CSE of $\pi^1$Gru using three different antenna configurations (see Sect. \ref{obs}). In the present work we present all morphological features identified in the ALMA data. In particular, we focus on the extended CO emission and on the more compact HCN and SiO emission. The data exhibit a prominent, finely resolved, inclined equatorial density enhancement with an embedded clockwise spiral that shows rotation near the stellar position. At large velocities, the contours of a high-velocity hourglass are discerned. This observational paper is the first in a series, in which follow-up papers will involve the modelling of the primary morphological features in the wind of $\pi^1$Gru using smoothed-particle hydrodynamics (SPH) and three-dimensional radiative transfer.
	
	The paper is organised as follows. We present a synthesis of the observation technicalities and reduction procedures in Sect. \ref{obs}. In Sect. \ref{continuum}, we analyse the continuum. This is followed by an analysis of the CO, HCN and SiO emission in Sect. \ref{molobs}. Subsequently, peculiarities in the data that deserve more attention are discussed in Sect. \ref{discus}. Finally, we present a summary of the results in Sect. \ref{summ}.
	
	\section{Data acquisition and reduction} \label{obs}
	
	$\pi^1$Gru was observed by ALMA between October 2018 and July 2019 using three different antenna array configurations.  For more details of observations and data processing see Gottlieb et al. \emph{in prep.}, Sects. 2 and 3. The synthesised beam size and the maximum recoverable scale (MRS) are given for the continuum, and vary slightly with frequency for spectral cubes. The data were combined with equal weight given to all configurations. We used multiscale CLEAN to produce the image cubes, with more weight given to large scales. When making image cubes for HCN and SiO a visibility plane taper equivalent to 0.02'' was applied to reduce CLEAN instabilities.  For CO, a taper of 0.17'' was used to further improve sensitivity to low brightness temperature features. The total image sizes were set to the smaller of: (i) the 0.2 sensitivity level of the primary beam, with a radius of $\sim$20", or (ii) the region encompassing all the detected emission. The general and continuum parameters are given in Table \ref{GENERAL} and the line cube details are given in Table \ref{LINE}.
	
	\begin{table}
		\caption{Summary of general and continuum data specifications.}
		\centering          
		\label{GENERAL}
		\begin{tabular}{llll}
			\hline\hline
			\noalign{\smallskip}
			Config. & Beam                  & Continuum & MRS \\
			& (mas$\times$mas,$^\circ$) & rms (mJy) & (arcsec)\\
			\noalign{\smallskip}
			\hline    
			\noalign{\smallskip}
			extended & 19$\times$19, 60    & 0.015 & 0.4 \\
			\noalign{\smallskip}
			mid      & 248$\times$235, 30  & 0.040 & 3.9 \\
			\noalign{\smallskip}
			compact  & 866$\times$774, -86 & 0.036 & 9.3 \\
			\noalign{\smallskip}
			combined & 25$\times$23, 44    & 0.010 & 9.3 \\
			\hline
		\end{tabular}
	\end{table}
	
	\begin{table*}
		\caption{Summary of the line cube data specifications.}
		\centering          
		\label{LINE}
		\begin{tabular}{lllclc}
			\hline\hline
			\noalign{\smallskip}
			Molecule & Transition & Rest freq. & Chan. width & Chan. rms               & Combined beam            \\
			&            & (GHz)      & ($\kms$)    & ext-mid-comp-comb (mJy) & (mas$\times$mas,$^\circ$) \\
			\noalign{\smallskip}
			\hline    
			\noalign{\smallskip}
			$^{12}$CO & $J$=2$-$1 & 230.5380 & 1.3 & 1.0\ --\ 3.5\ --\ 3.0\ --\ 1.1 & $292\times278$, -22 \\
			\noalign{\smallskip}
			HCN & $J$=3$-$2 & 265.8864 & 1.1 & 1.2\ --\ 3.0\ --\ N/A\ --\ 1.5 & $53\times46$, 61 \\ 
			\noalign{\smallskip}
			SiO & $J$=5$-$4 & 217.1050 & 1.4 & 1.1\ --\ 3.5\ --\ 2.9\ --\ 1.0 & $51\times43$, 17 \\
			\hline
		\end{tabular}
		\vspace{1ex}\\
		\begin{flushleft}
			\textbf{Note}: All lines are in $\varv = 0$. 
		\end{flushleft}
	\end{table*}

    \begin{figure}[]
        \centering
        \includegraphics[width=8cm]{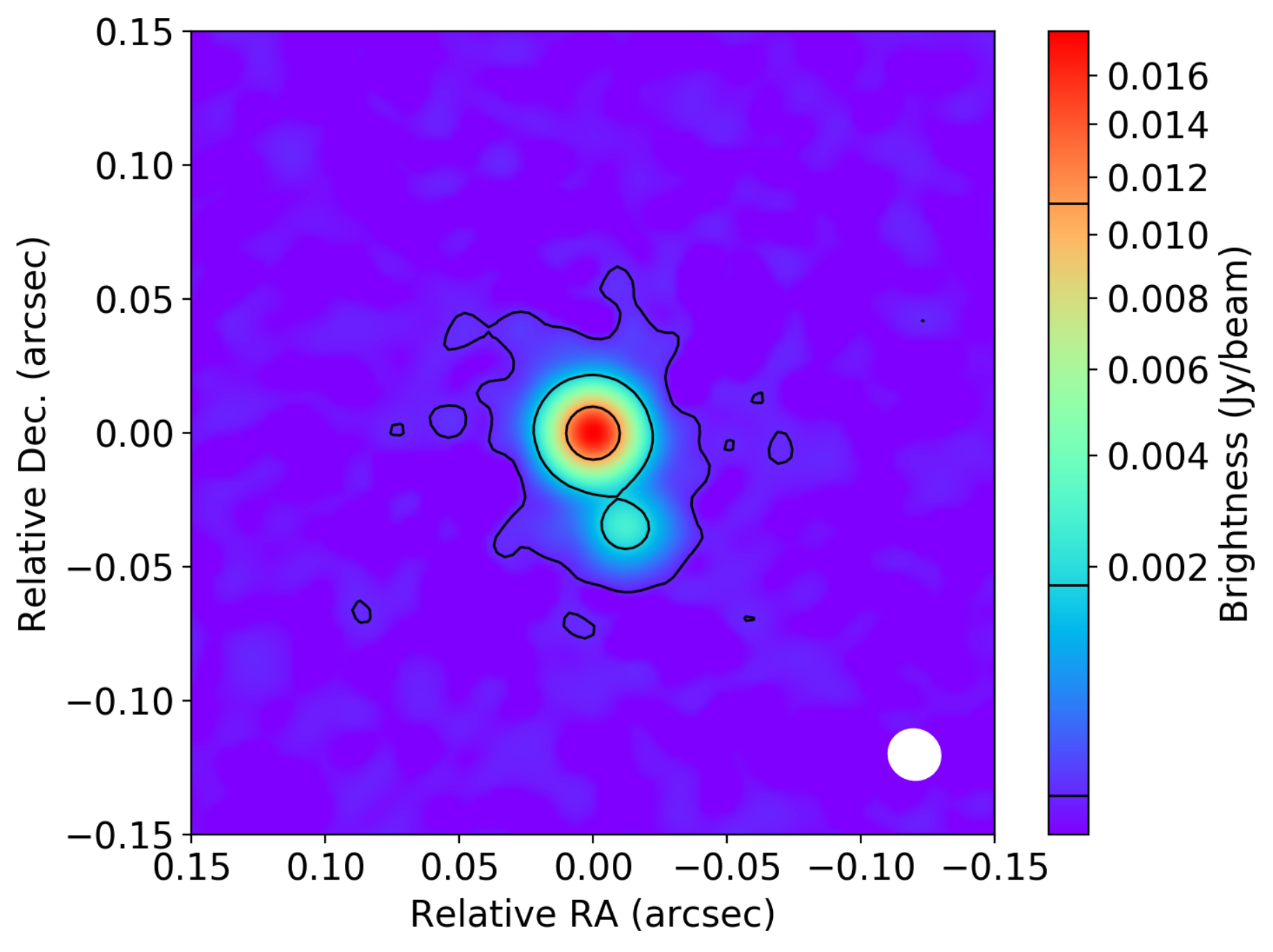}
        \caption{Continuum emission of the $\pi^1$Gru system, as observed by the extended 12m configuration, at 211 -- 275~GHz. Contours are drawn at 3, 120, and 768 times the continuum rms noise value (1.5 $\times {\rm 10}^{\rm -5}$ Jy/beam). The ALMA beam size is shown in the bottom right corner. The continuum shows a second peak to the south-west of the central emission.}
        \label{cont}
    \end{figure}

    \begin{figure}[]
        \centering
        \includegraphics[width=8.5cm]{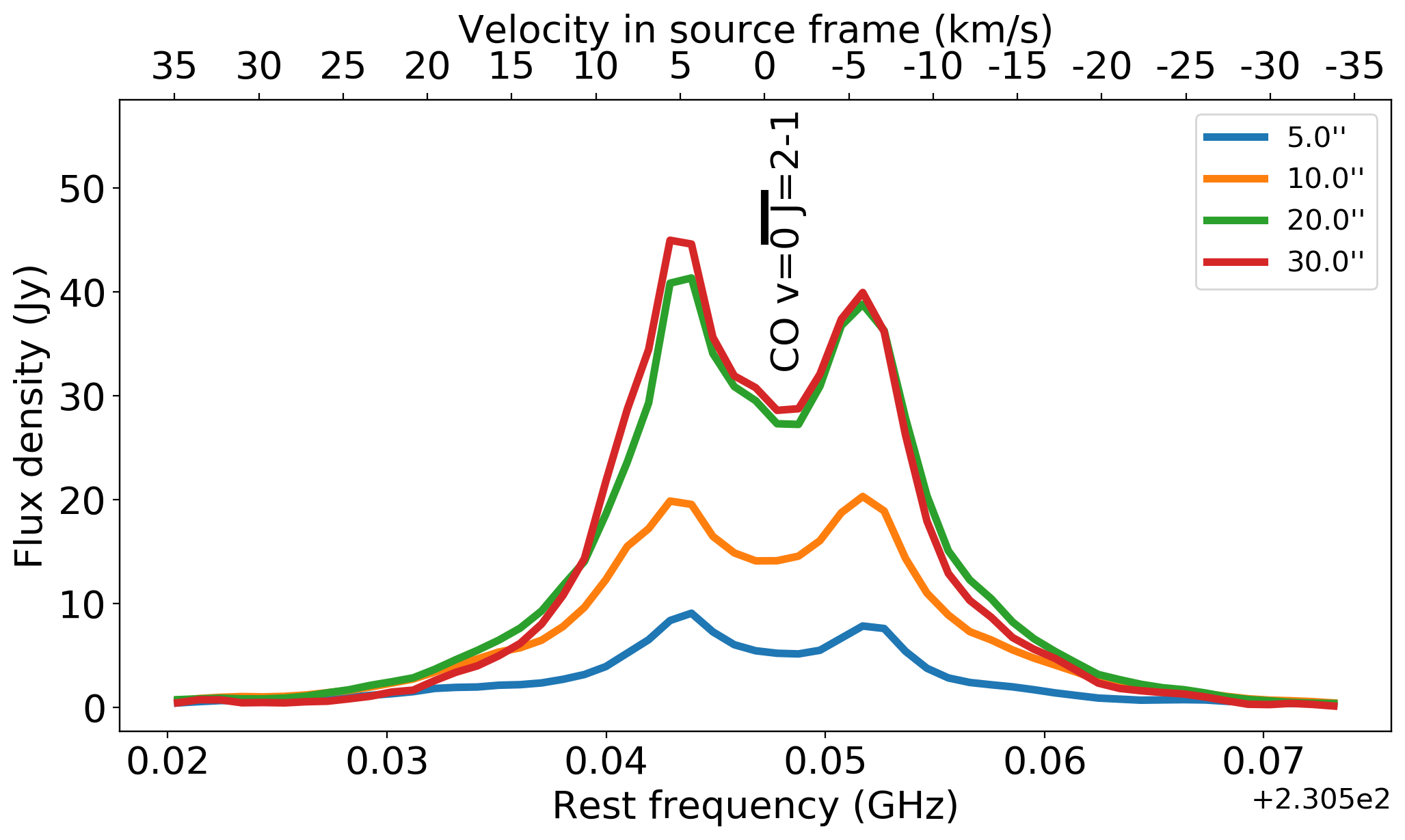}
        \caption{Line spectrum of the CO v=0 $J=\ $2$-$1 spectral line, for different aperture diameters. The frequency-axis is adjusted to the stellar velocity.}
        \label{COline}
    \end{figure}

    \begin{figure}[]
        \centering
        \includegraphics[width=7.0cm]{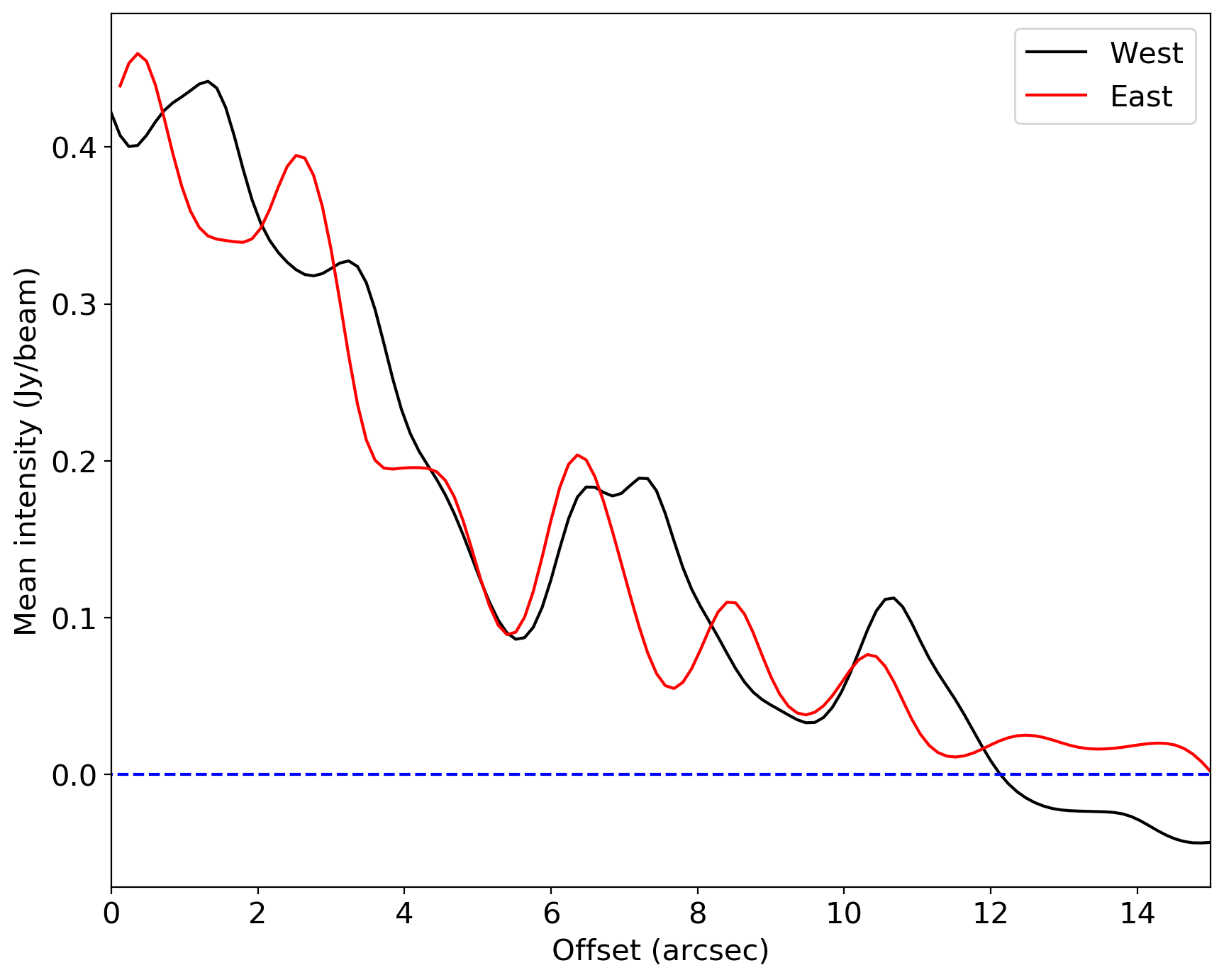}
        \caption{Radial profile of the channel at stellar velocity of the CO cube shown in Fig. \ref{COchan}, along the axis with PA=96$^\circ$, and a width of 1 compact configuration beam. The 3$\times \sigma_{\rm rms}$ noise level is smaller than the line thickness. Offset is measured with respect to the AGB star coordinates.}
        \label{COradial}
    \end{figure}

\begin{figure*}[]
        \centering
        \includegraphics[width=7cm]{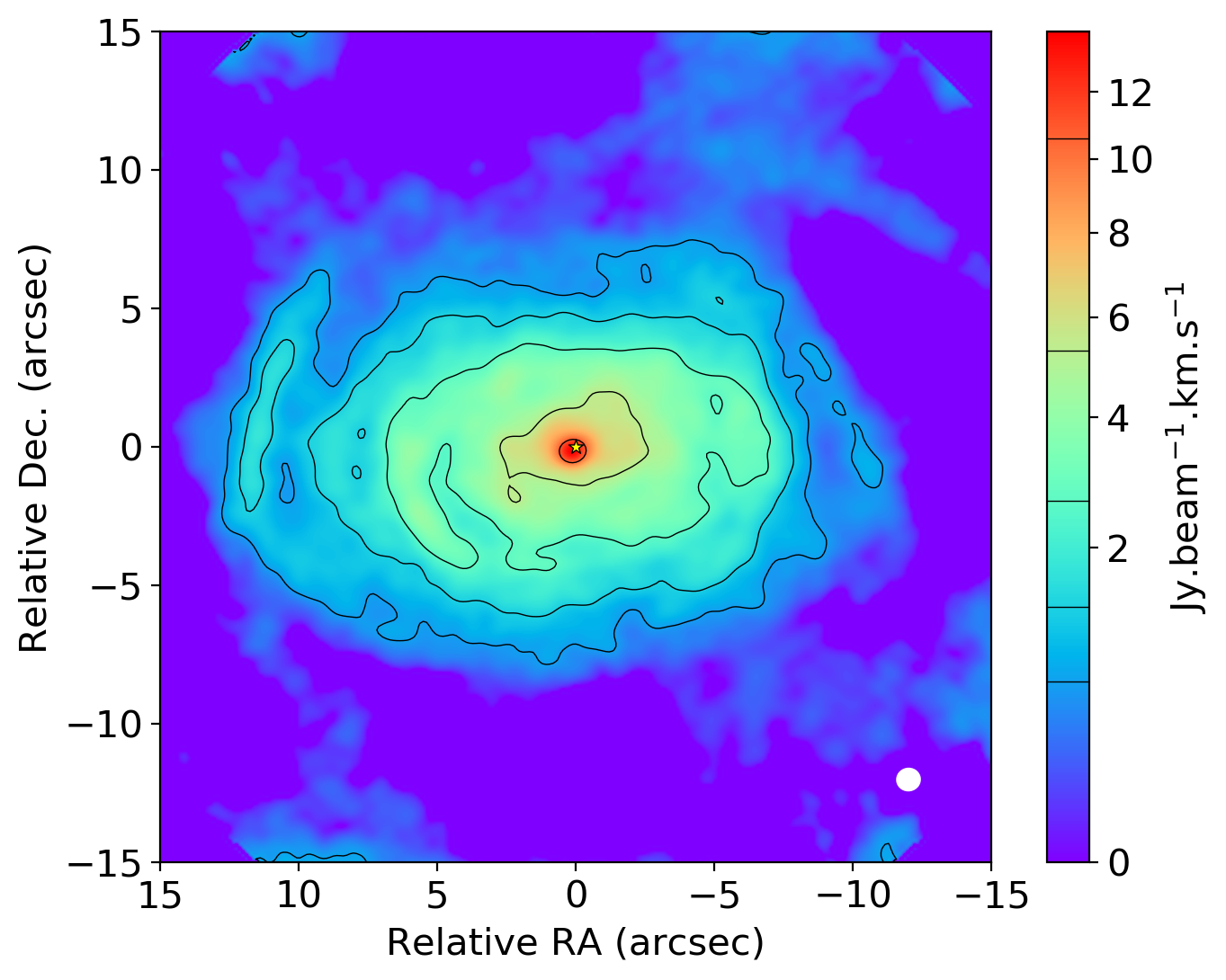}
        \includegraphics[width=7cm]{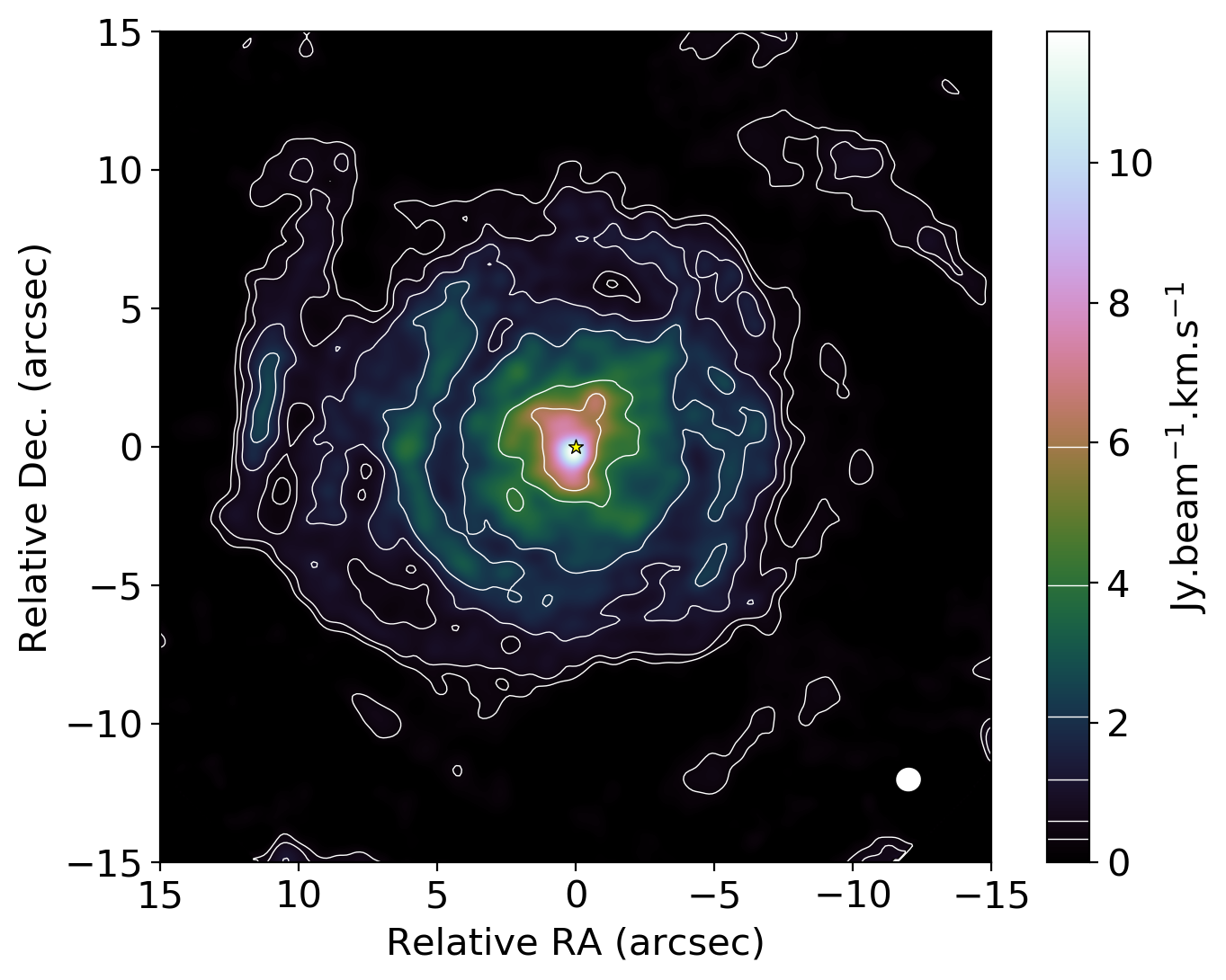}
        \caption{\emph{Left:} moment0 map of the CO compact configuration emission in the channel maps shown in Fig. \ref{COchan}. Contours are drawn at 3, 6, 12, 24, 48, and 96 times the rms noise value in the spectral region of the bandpass without detectable line emission ($\sigma_{\rm rms}$ = 3.6$\times {\rm 10}^{\rm -3}$ Jy/beam). The continuum peak position is indicated by the yellow star symbol. \emph{Right:} Velocity-integrated emission for the cherry-picked channels at $\sim$ 11.5, 5, -2, and -8.5~$\kms$, selected and displayed as to highlight the spiral-like features in the emission.}
        \label{COmom0}
\end{figure*}
	
	\section{ALMA continuum emission} \label{continuum}
	
	The continuum emission is shown in Fig. \ref{cont}. It consists of two components, one bright slightly resolved component to the north and another, completely unresolved, much dimmer component to the south-west. The components are separated by approximately 0.038'' (or about 6.25~au assuming a distance of 163 pc) in the plane of the sky. We shall henceforth refer to the brighter component as the primary, and the dimmer component as the secondary. The two components are separated at the 120$\times \sigma_{\rm rms}$ level. The brightness peak of the primary is about 17~mJy/beam, and it has a total integrated flux density of 25~mJy within the northern loop of the 120$\times \sigma_{\rm rms}$ contour. The diameter of this loop is approximately 0.045''. The secondary has a brightness peak of 2.7~mJy/beam, and a total integrated flux density within the southern loop of the 120$\times \sigma_{\rm rms}$ contour of 1.3~mJy. The size of the south loop has a diameter of approximately 0.02''. 
	
	Assuming the primary component represents the AGB star of the system, for which the surface temperature and luminosity have been previously measured (see Sect. \ref{intro}, though not accounting for the dual nature of the central source), we can estimate its contribution to the total continuum flux (by assuming black-body emission in a bandwidth 1.6 GHz around the CO rest frequency of 230.5~GHz) to be 26.5~mJy, or approximately the total flux within the north loop of the 120$\times \sigma_{\rm rms}$ contour. Even though this estimation assumes no attenuation by circumstellar dust, when considering the uncertainties of $\sim$20\% on the measurement set and even higher uncertainties on the previously determined stellar parameters, it is safe to assume that the primary component indeed fully represents the AGB star.
	
	The brightness peak of the secondary source in the continuum is located at a projected distance of $\sim$6~au away from the central star, and could hence be the predicted companion \citep{Sahai1992,Mayer2014} that is shaping the wind, or a local environment created by a companion (see Sect. \ref{dynamics}). The flux ratio between the secondary and the primary is about 5\% at $\sim$230~GHz. Assuming that all energy is of thermal origin and that the two sources are radiating a Planck field in the Rayleigh-Jeans regime from a surface of radius R, then it can be shown that
	\begin{equation}
	R_{\rm sec} = 78\rso \left( \frac{3100K}{T_{\rm sec}}\right)^{\frac{1}{2}},
	\end{equation}
	where $R_{\rm sec}$ is the surface radius of the secondary object, and $T_{\rm sec}$ is its temperature. The above calculation strongly suggest that the secondary has either a comparatively large surface area, or a high temperature, or both. This points towards the presence of a high degree of accumulated, very hot dust. At this stage we are unable to conclusively distinguish between a compact dense clump of hot dust, or an accretion structure surrounding a potential (stellar) companion. However, analysis of the dynamics of the structures found in the wind of $\pi^1$Gru (see Sects. \ref{maserdiscus} and \ref{dynamics}) strongly suggest that the secondary continuum peak is indeed located at the expected companion position, which provides support for the latter scenario.
	
	In addition, full analysis of the properties of the ALMA continuum, in combination with SPHERE/ZIMPOL data, which confirms that the feature is indeed a companion, is currently in preparation (Montarg\`{e}s et al., \emph{in prep.}). We therefore defer all further examination and discussion on the nature of the secondary component to this work.
	
	\section{ALMA molecular emission} \label{molobs}
	
	In this section we describe the observational results for the brightest spectral lines in the {\sc Atomium} dataset that trace the morphology of $\pi^1$Gru: CO v=0 $J=\ $2$-$1 \citep{1997JMoSp.184..468W}, and, for the first time, HCN v=0 $J=\ $3$-$2 \citep{2002ZNatA..57..669A} and SiO v=0 $J=\ $5$-$4 \citep{2013JPCA..11713843M} (See Table \ref{LINE}). All spectrally resolved figures and mentions of blue-shift and red-shift are made with respect to the stellar radial velocity $v_*$, which is -12~$\kms$ with respect to the local standard of rest. For optimal appraisal of the fine details in the data figures we recommend that all images presented in this paper are displayed on screen. For all resolved maps north is up, and east is left.

	\subsection{CO $J=\ $2$-$1 emission: Large-scale clockwise spiral} \label{COobs}
	
	The description of the global morphological properties of the CSE based on the CO emission requires a uniform representation of the data that brings out features with a low surface brightness. We therefore focus this initial analysis on the compact 12m main array data. The finer morphological details will be described using the combined (compact, mid and extended) 12m ALMA array configurations datacube.
	
	Fig. \ref{COline} shows the spectral line of the observed CO $J=\ $2$-$1, produced from the compact configuration. It exhibits a `double-horned' line shape, with peaks around $v_*\pm {\rm 5}\ \kms$, and extremely broad line wings, extending to projected velocities of at least $v_*\pm {\rm 30}\ \kms$, with an integrated flux of about $\sim$820 Jy$\kms$ for the largest aperture. 
	
	Use of the combined 12m array data alone (MRS = 9.3'') implies that the data is missing flux. To estimate this missing flux we compare the integrated flux of the current CO line with single-dish data in the literature. \citet{Winters2003} show in their Fig. 23 a CO $J=\ $2$-$1 emission line observed with the 15m diameter Swedish-ESO Submillimeter Telescope (SEST), for which we have calculated its rescaled integrated flux to $\sim$1140 Jy$\kms$. \citet{Doan2017} show APEX data from 2005 for which we calculated an integrated flux of $\sim$3080 Jy$\kms$. This means that between 27\% and 72\% of the diffuse, larger-scale emission is missing from the data. Nevertheless, since all analyses below are based on emission features with typical dimensions that are much smaller than the MRS of the configuration, we do not expect this diffuse emission to impact any of our conclusions. 

	
	The channel maps in Fig. \ref{COchan} show the distribution of emission in the $\pm {\rm 18}\ \kms$ velocity range around the systemic stellar velocity for the compact and combined 12m main array data. The emission dominates the northern (southern) quadrants in the red- (blue-)shifted projected velocities, and shows a near-perfect alignment with the continuum brightness peak position at stellar velocity. We summarise the general properties of the global nebula in Sect. \ref{A.CO} in the appendix. These are in agreement with the results by \citet{Doan2020}, so we refrain from repeating them here.
	
	The emission is not smoothly decaying with radius, contrary to cases like L$_{\rm 2}$ Pup \citep{Homan2017} or R Dor \citep{Homan2018}. Rather, the emission appears to show a multitude of second-order arc-like enhancements. The systematic changes of the location of these emission enhancements as a function of radial velocity reveals spectral continuity, which suggests that they represent a coherent entity. The most straightforward explanation for such globally linked arcs would be that the EDE of $\pi^1$Gru possesses an embedded spiral, as also first proposed by \citet{Doan2020}. 
	
	To analyse the positions of these arcs with respect to each other we construct a radial profile of the central channel of the CO data along the symmetry axis of the stereogram (see Fig. \ref{COstereo}), averaged over 0.8'', the resolution of the ALMA compact configuration images. This radial profile is shown in Fig. \ref{COradial}. In the inner 5'', the peaks appear somewhat periodically, while also showing a systematic offset between the ones to the east and west: the peak positions of the western (eastern) emission coinciding with the troughs of the eastern (western) emission, as expected for a spiral. The spiral arms are nearly equidistant with a spacing of $\sim$2'' ($\sim$325~au at a distance of 163 pc). The width of these arcs is approximately 1'' at all radii, and the peaks do not appear to show a consistent brightness contrast with respect to the troughs. 
	
	Beyond 5'' the emission peaks in the eastern profile remain somewhat equidistant, while in the western profile this trend is lost. Around 7'' the western profile exhibits a double peak, followed by a peak around 11'': the distance between these peaks is almost twice that of the peak spacing of the eastern profile. This indicates that the spiral may be undergoing geometrical changes as it moves away from the star. We elaborate on these curiosities in Sect. \ref{spiraldiscus}
	
	
	In many cases the moment0 map (the velocity averaged emission distribution over the entire domain shown in Fig. \ref{COchan}) is a useful tool to evaluate the degree of spectral and spatial correlation of emission patterns throughout the entire spectral domain of the datacube. However, as can be seen in the left panel of Fig. \ref{COmom0}, other than some arc-like features, it does not show any signs of spiral-like correlated emission. This could be caused by a combination of a relatively large system inclination ($\sim$40$^\circ$ as estimated by \citet{Doan2017}), and a relatively large equatorial thickness angle. Combined, this would cause the lower-emission gaps between the spiral arms to be covered by the emission of the arms at larger radii. Hence, we construct a similar map from carefully chosen velocity bins to mitigate the effect. Shown in the right panels of Fig. \ref{COmom0} is the velocity averaged map for the velocity bins at 11.5, 5, -2, and -8.5~$\kms$, which we shall refer to as the CPmom0 map (CP = cherry pick). This map highlights the consecutive bright, curved arcs of emission that indeed appear to trace the outline of a clockwise spiral, which is opposite to the conclusion drawn by \citet{Doan2020}.
	
	Finally, we also recover the high-velocity signals that were first detected by \citet{Sahai1992}, and later imaged by \citet{Doan2020}, as shown in Fig. \ref{COwing}. The properties of the detected features are in complete agreement with the findings of \citet{Doan2020}, and, to avoid repetition, we summarise our results in Sect. \ref{A.hg} in the appendix.
	
	\subsection{HCN $J=\ $3$-$2 emission: Innermost spiral windings} \label{HCNobs}
	
	\begin{figure}[]
		\centering
		\includegraphics[width=8.5cm]{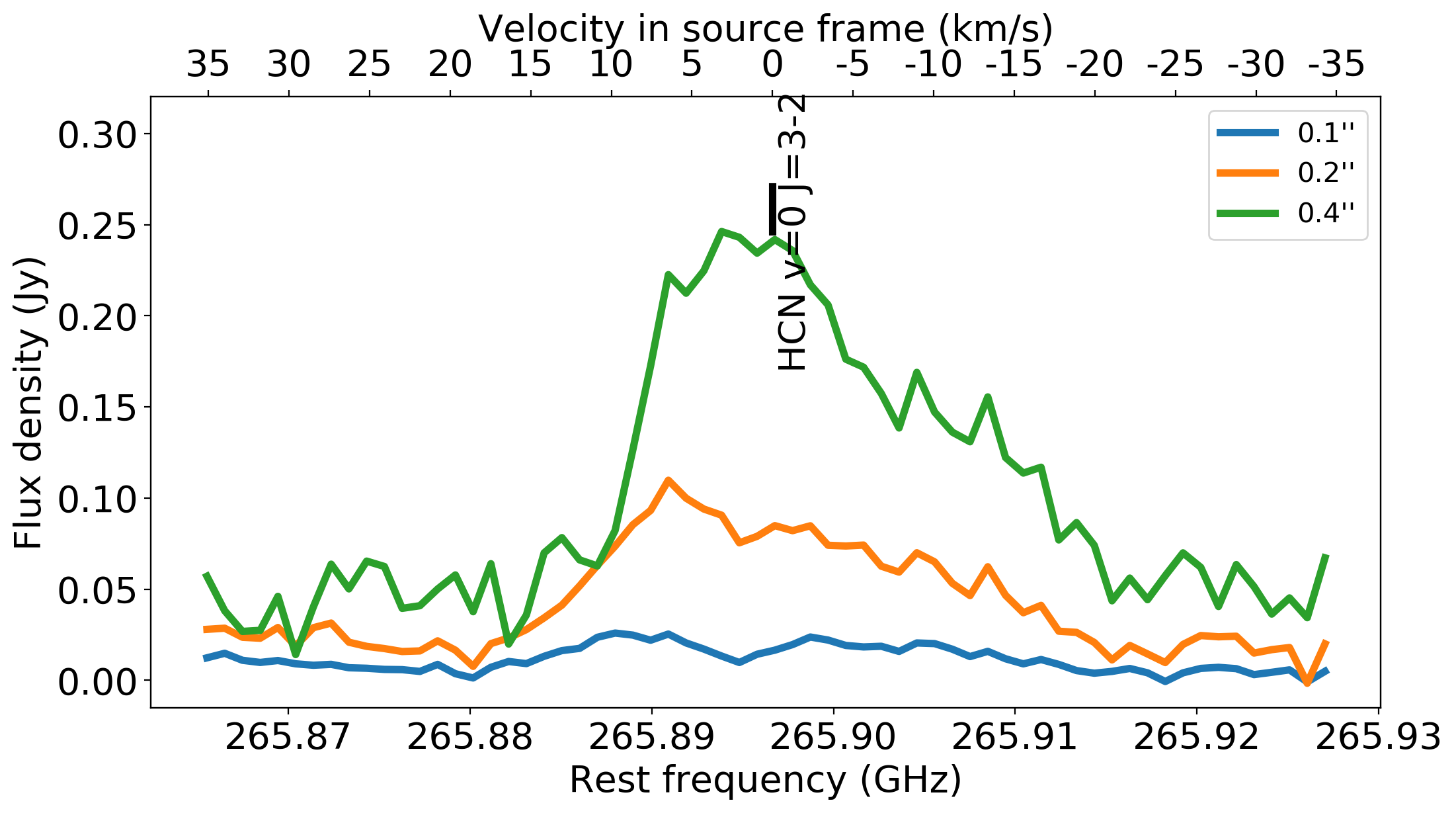}
		\caption{Line spectrum of the HCN v=0 $J=\ $3$-$2 spectral line for the combined dataset, for different aperture diameters. The frequency-axis is adjusted to the stellar velocity.}
		\label{HCNline}
	\end{figure}
	
	\begin{figure*}[]
		\centering
		\includegraphics[width=16cm]{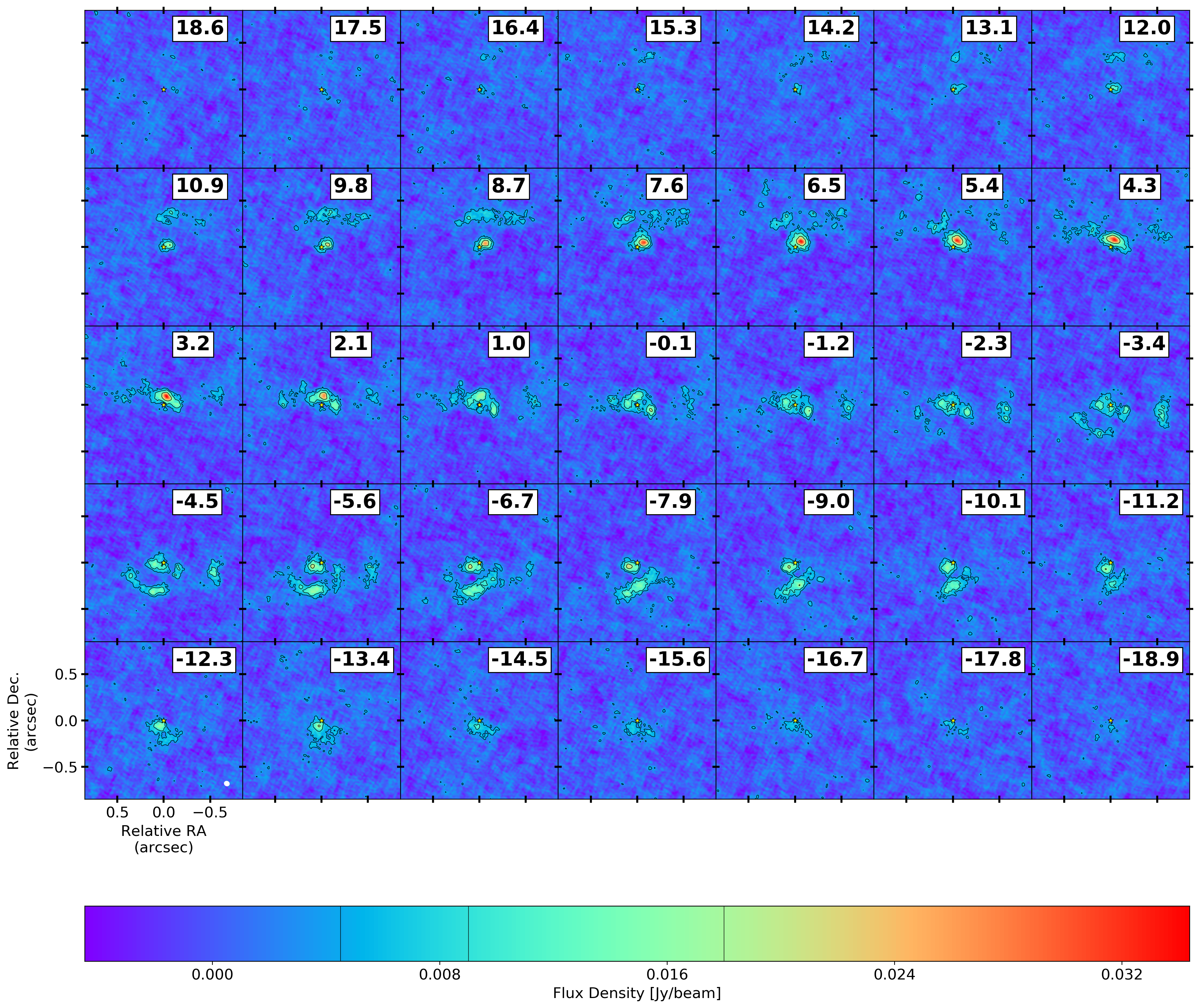}
		\caption{Channel maps showing the resolved emission of the HCN v=0 $J=\ $3$-$2 line in the $\pm \sim$18~$\kms$ velocity range, for the combined dataset. The labelled velocities have been corrected for $v_{*}$=-12~$\kms$. Contours are drawn at 3, 6, 12, 24, 48, 96, and 192$\times \sigma_{\rm rms}$ (= 1.5$\times {\rm 10}^{\rm -3}$ Jy/beam). Length scales and ALMA beam (0.053''$\times$0.046'') are indicated in the bottom left panel. The maps are centred on the continuum peak position, which is indicated by the yellow star symbol.}
		\label{HCNchan}
	\end{figure*}
	
	\begin{figure*}[]
		\centering
		\includegraphics[width=8.5cm]{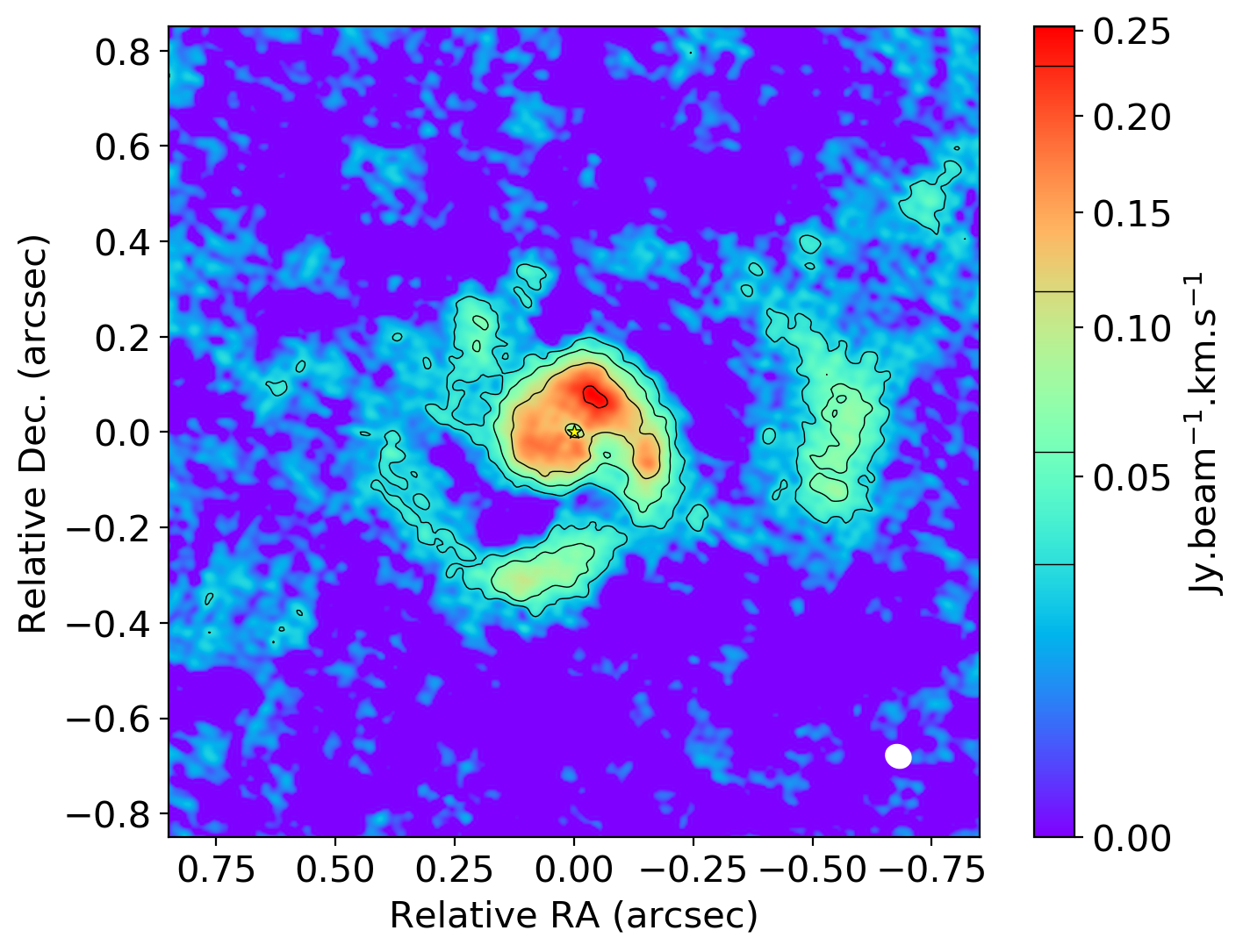}
		\includegraphics[width=8.8cm]{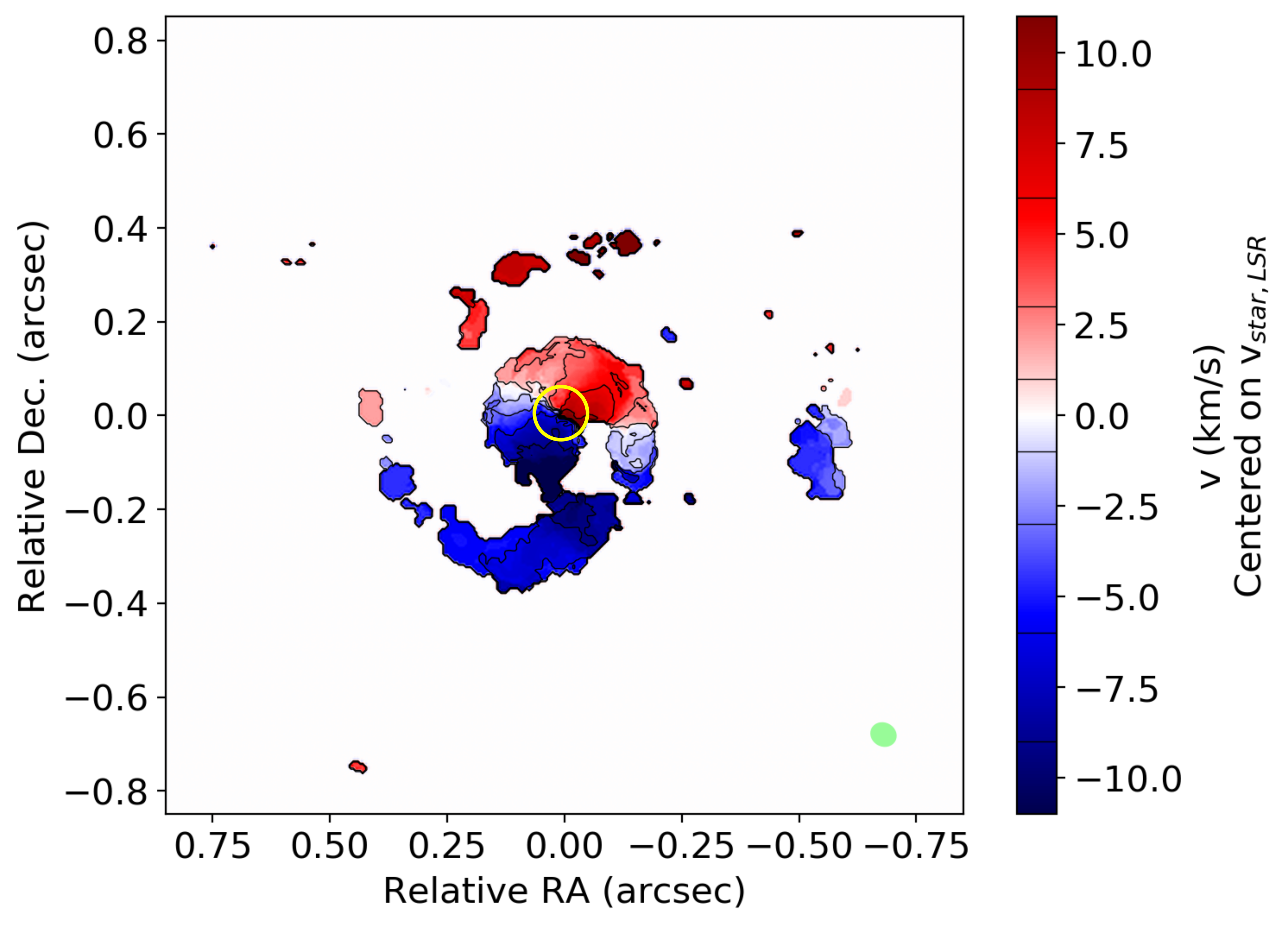}
		\caption{\emph{Left:} moment0 map of the emission in the channel maps shown in Fig. \ref{HCNchan}. Contours are drawn at 3, 6, 12, 24, 48, and 96 times the rms noise value in the spectral region of the bandpass without detectable line emission ($\sigma_{\rm rms}$ = 1.5$\times {\rm 10}^{\rm -3}$ Jy/beam). The continuum peak position is indicated by the yellow star symbol. \emph{Right:} moment1 map of the same emission. Contours are drawn at $\pm$1, $\pm$3, $\pm$6 and $\pm$9~$\kms$. The zero-velocity strip separating the blue- and red- shifted emission inside the small central yellow circle has a different PA than that of the zero-velocity strip separating the blue- and red- shifted emission CO emission	(see Fig. \ref{COmom1}).}
		\label{HCNmom}
	\end{figure*}
	
	The HCN emission in the dataset is confined to the inner 1.5'' zone (see Fig. \ref{HCNchan}), thus focusing precisely on the region where the first spiral winding is detected in the CO emission (see Fig. \ref{COradial}). Fig. \ref{HCNline} shows the spectral line of the HCN emission, for different aperture diameters. The line is highly asymmetric with a steep red-shifted line wing, and a broad blue-shifted line wing. The line peaks at a spectral offset of $\sim v_*$+4~$\kms$, indicating that the local emission conditions deviate highly from sphericity. The line does not display any bright, narrow features that one would expect for maser emission.
	
	We show the spatially resolved HCN emission in the channel maps in Fig. \ref{HCNchan}. In the red (blue)-shifted portion of the line wings, emission can be seen in arc-like filaments to the north (south) of the AGB star. As speeds approach the stellar velocity, the emission becomes stronger, the filaments split into two emission clumps that align themselves horizontally with the AGB star. This spectral progression strongly resembles what is seen in the CO channel maps, and indicates that the large-scale EDE morphology as manifested in CO emission persists down to sub-arcsecond length scales. This is confirmed by the moment0 and moment1 maps shown in Fig. \ref{HCNmom}, which reveal a clockwise spiral consisting of 2 complete revolutions. Hence, the HCN data confirm that the emission patterns analysed in the Sect. \ref{COobs} indeed follow a spiral pattern that originates near the location of the AGB star. This result stands in contrast with the interpretation proposed by \citet{Doan2020}. The HCN emission patterns now unambiguously demonstrate that the spiral is in fact clockwise, and shows no signs of perturbations brought about by a companion in an eccentric orbit.
	
	The HCN emission never actually overlaps with the AGB star. This implies that the detected HCN emission is solely confined to the spiral arm. It is known that the HCN molecule is a product of non-LTE chemistry in M- and S-type stars \citep{Gobrecht2016,VandeSande2020}. Hence, the fact that the HCN emission is \emph{only} detected in large quantities in the spiral implies that the conditions associated with the spiral and its formation can, and probably does, induce or enhance non-LTE chemical processes.
	
	The HCN spiral has a consistent thickness of 0.12'' and a spacing between the windings of 0.3''. And whereas most of the emission in the HCN spiral appears rather homogeneously distributed at intensities around 1.2$\times$10$^{\rm -2}$ Jy/beam, the HCN spiral exhibits an unusually bright clump of emission that persists over the 3 to 10~$\kms$ velocity range. This clump grazes the AGB star to the north-west. It has a length (measured along its longest dimension) of 0.3'', and a width of 0.17'', and it has a brightness peak position at a distance of approximately 0.9'' from the AGB star. This clump which, with its brightness peak of 3.3$\times$10$^{\rm -2}$ Jy/beam, is approximately 3 times brighter than the rest of the spiral, is also visible in the moment0 map in Fig. \ref{HCNmom}.
	
	In the moment1 map, shown in the right panel of Fig. \ref{HCNmom}, the north-to-south spectral separation also observed in the CO emission (see Fig. \ref{COmom1}) is clearly visible. But, the HCN emission reveals a substantial deviation from this trend below 0.1'' (inside the yellow circle in the right panel of Fig. \ref{HCNmom}). The position-angle (PA, measured from north to east) of the strip representing the gas with projected speeds close to the systemic velocity in the inner wind is $\sim$55$^\circ$, nearly 45$^\circ$ rotated with respect to the previously deduced orientation of this strip at larger scales (e.g. Fig. \ref{COmom1}). This trend is also recovered from the SiO emission discussed in the next section, and will be addressed in Sect. \ref{rotdiscus}.
	
	\subsection{SiO $J=\ $5$-$4 emission: Gas in rotation} \label{SiOobs}
	
	\begin{figure}[]
		\centering
		\includegraphics[width=8.5cm]{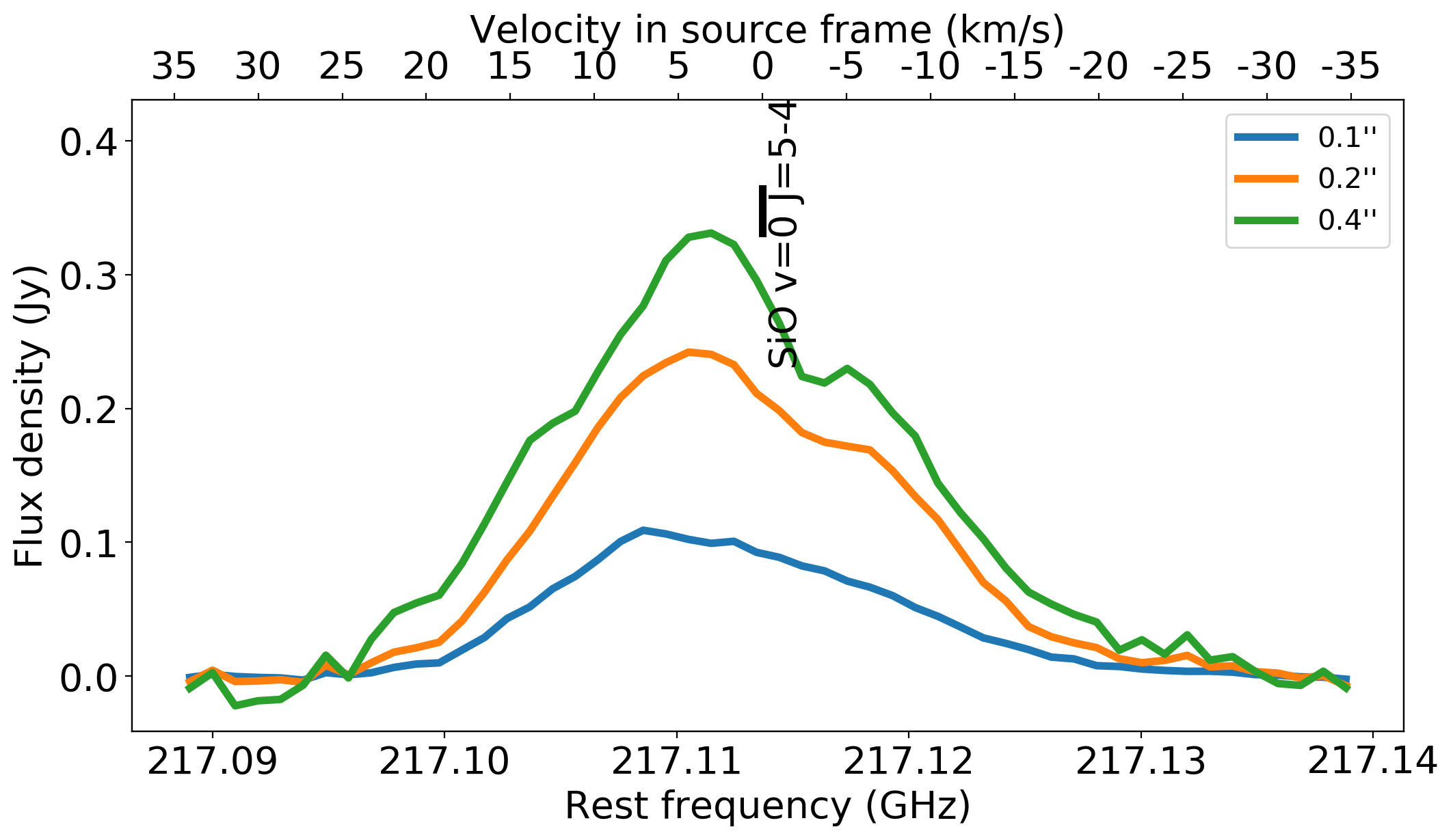}
		\caption{Line spectrum of the SiO v=0 $J=\ $5$-$4 spectral line for the combined dataset, for different aperture diameters. The frequency-axis is adjusted to the stellar velocity.}
		\label{SiOline}
	\end{figure}
	
	\begin{figure*}[]
		\centering
		\includegraphics[width=16cm]{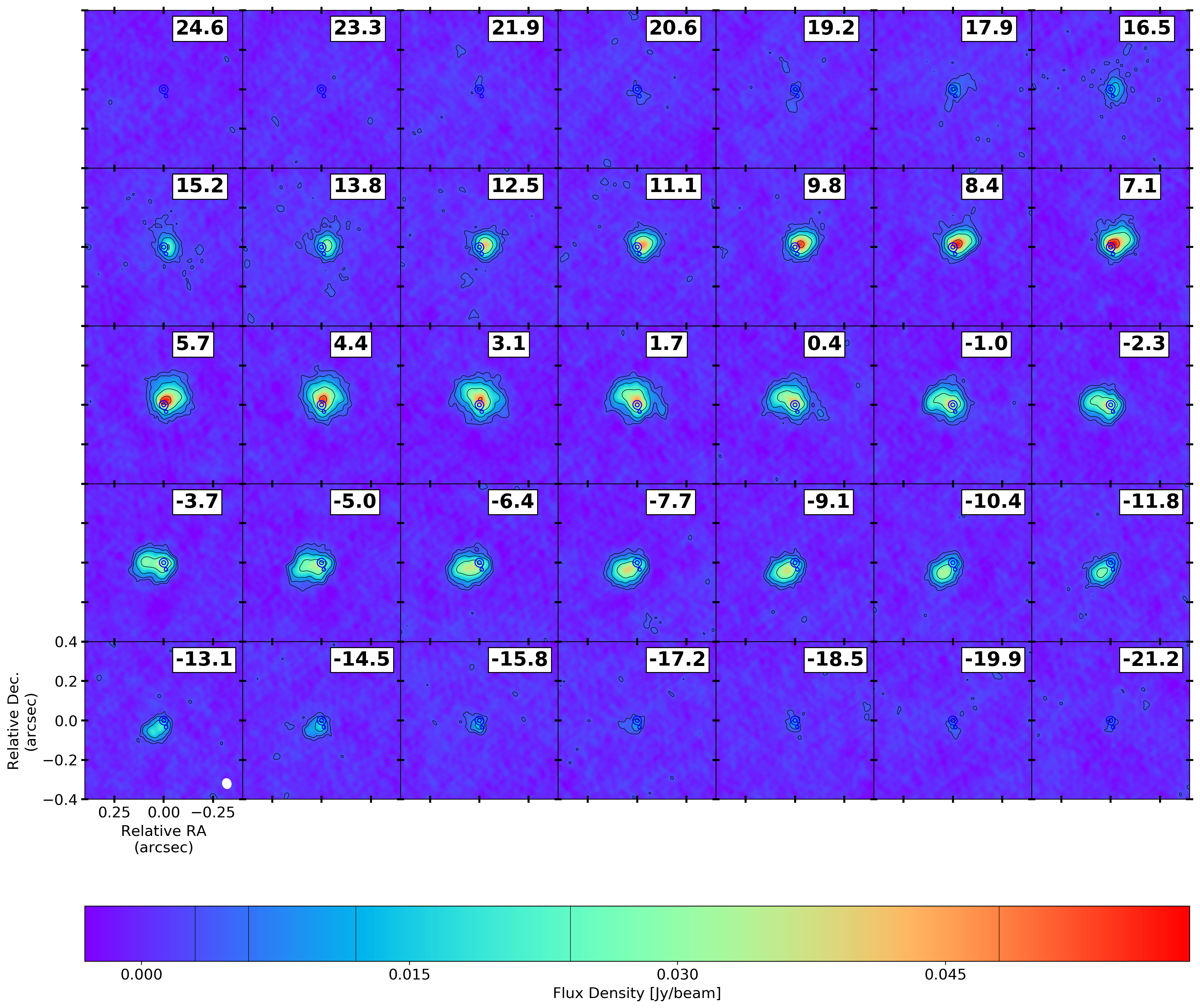}
		\caption{Channel maps showing the resolved emission of the SiO v=0 $J=\ $5$-$4 line in the $\pm \sim$23~$\kms$ velocity range, for the combined dataset. The labelled velocities have been corrected for $v_{*}$=-12~$\kms$. Contours are drawn at 3, 6, 12, 24, 48, 96, and 192$\times \sigma_{\rm rms}$ (= 1.0$\times {\rm 10}^{\rm -3}$ Jy/beam). Length scales and ALMA beam (0.051''$\times$0.043'') are indicated in the bottom left panel. The maps are centred on the continuum peak position. The two highest continuum contours in Fig. \ref{cont} are shown in blue.}
		\label{SiOchan}
	\end{figure*}
	
	\begin{figure*}[]
		\centering
		\includegraphics[width=8.5cm]{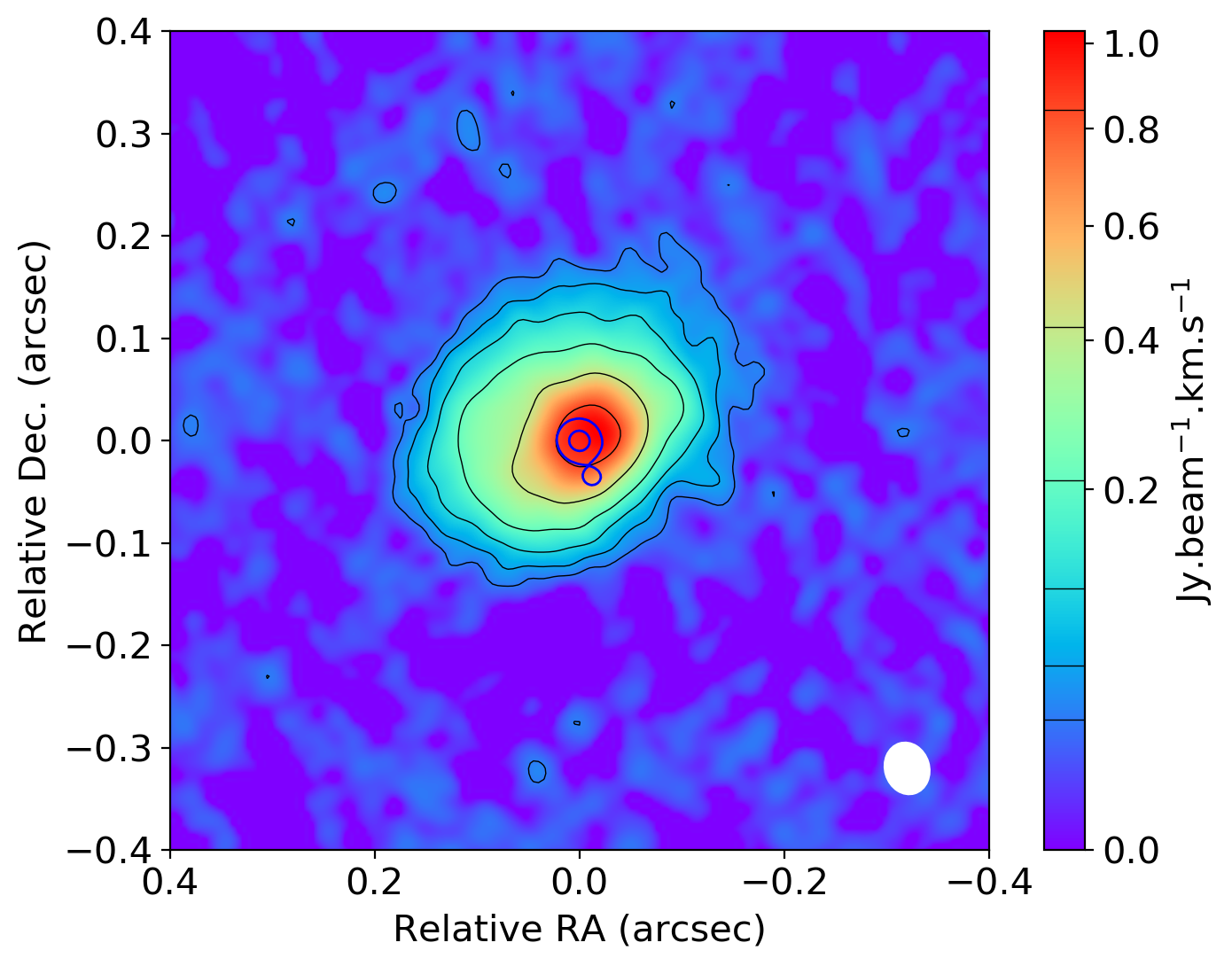}
		\includegraphics[width=8.8cm]{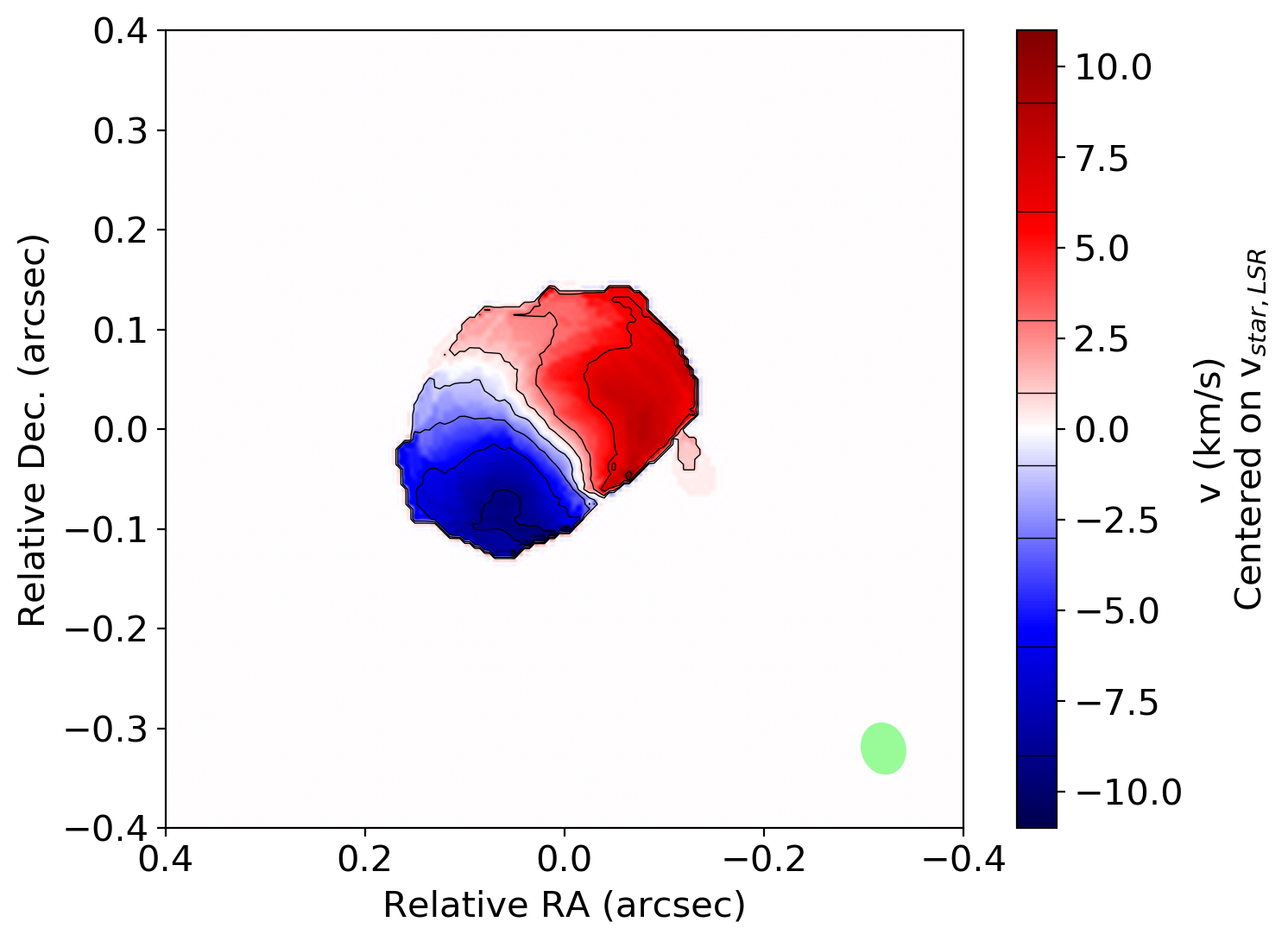}
		\caption{\emph{Left:} moment0 map of the emission in the channel maps shown in Fig. \ref{SiOchan}. Contours are drawn at 3, 6, 12, 24, 48, and 96 times the rms noise value in the spectral region of the bandpass without detectable line emission ($\sigma_{\rm rms}$ = 1.0$\times {\rm 10}^{\rm -3}$ Jy/beam). The two highest continuum contours in Fig. \ref{cont} are shown in blue.  \emph{Right:} moment1 map of the same emission.}
		\label{SiOmom}
	\end{figure*}
	
	The spectral line of the SiO v=0 $J=\ $5$-$4 emission is shown in Fig. \ref{SiOline}. It has a nearly-triangular line shape, with a symmetric velocity extent from about ${\rm -25}\ \kms$ to about ${\rm 25}\ \kms$. It does not show any spike-like features reminiscent of maser-activity. The only observed deviation is a small asymmetric dip around -3~$\kms$. Under the simple assumption of a spherical wind, such triangular line shapes are known to be tracers of the acceleration region of the wind \citep{Bujarrabal1986,Bujarrabal1989}. However, the analysis of the CO and HCN emission in the respective Sects. \ref{COobs} and \ref{HCNobs} has revealed that the nebula of $\pi^1$Gru is highly complex, with yet-to-be explained spectral behaviour of the gas at length scales comparable to the binary separation (see Sect. \ref{HCNobs}). Thus, the assumption of a smooth accelerating radial wind is probably insufficient to fully understand the origin of the line shape.
	
	The channel maps of the SiO emission for the combined dataset, shown in Fig. \ref{SiOchan}, display how the emission is strictly confined to regions that are at most 0.25'' away from the AGB star. With a photospheric radius for the AGB star of $\sim$0.02'' \citep{Paladini2018}, this means that the SiO probes the warm gas within the first ten stellar radii. Similar to the HCN emission, the evolution from the red- to the blue-shifted line wing of the emission closest to the AGB star manifests itself as what seems to be counter clockwise revolution from west to south-east around the AGB star. Though this behaviour is partly attributable to the presence of a spiral-shaped density enhancement surrounding the AGB star, the eastern (western) positioning of the highest blue-(red-)shifted emission is unexpected for a purely radial velocity field.
	
	Contrary to HCN, SiO is a stable and easy-to-produce product of LTE chemistry in oxygen-rich environments \citep{Gail2013b}. This implies that it is commonly found in relatively large abundances in the inner wind. This can also be seen in the moment0 map shown in the left panel of Fig. \ref{SiOmom}: the SiO emission fully encompasses the AGB star. Unlike CO, SiO possesses a relatively high electric dipole moment of $\sim$3.01~Debye \citep{Raymonda1970,Maroulis2000}. Thus, SiO tends to be more of a radiatively excitable molecule (unless affected by high optical depth effects). Close to the AGB star, where the stellar radiation field dominates, this makes a single SiO emission line an unreliable tracer for the local density landscape. However, the distribution and relatively high Einstein A coefficient of the current transition make the $J$=5-4 line an excellent tracer of the velocity field around the star, which we visualise using the moment1 map (Fig. \ref{SiOmom}, right panel). It reveals that the blue-shifted and red-shifted gas are separated by a region where the line-of-sight velocity is close to the systemic velocity, which manifests itself as a long, thin strip with a PA=50$^\circ$, just as detected for the HCN emission in Sect. \ref{HCNobs}. We discuss this further in the Sect. \ref{rotdiscus}. 
	
	\subsection{Smallest-scale SiO emission}
	
\begin{figure}[]
		\centering
		\includegraphics[width=8.5cm]{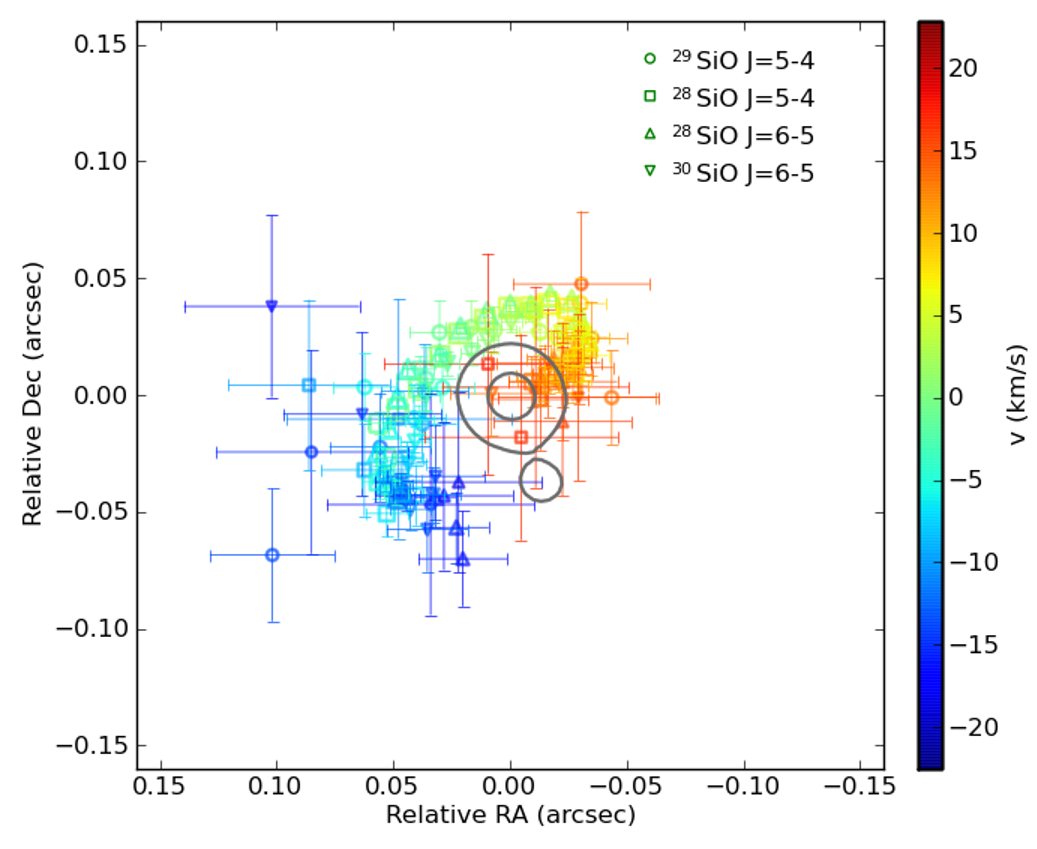}
		\caption{Position and velocity of the localised peak positions of the ground-state vibrational $^{\rm 28}$SiO,$^{\rm 29}$SiO $J=\ $5$-$4 and $^{\rm 28}$SiO, $^{\rm 30}$SiO $J=\ $6$-$5 emission lines of the $\pi^1$Gruis {\sc Atomium} mid-configuration dataset. Contours of the 120 and 768 times the continuum rms noise value (1.5 $\times {\rm 10}^{\rm -5}$ Jy/beam) are shown in grey.}
		\label{maser}
	\end{figure}

	The ground-state vibrational $^{\rm 29}$SiO $J=\ $5$-$4, $^{\rm 28}$SiO $J=\ $6$-$5, and $^{\rm 30}$SiO $J=\ $6$-$5 lines are within our bands \citep{Mollaaghababa1991}. Although unlikely to be strongly masing, they all show bright, compact emission within $\sim$0.1'' of the star. Though even more compact, the emission distribution of these lines is geometrically and spectrally nearly identical to the SiO v=0 $J=\ $5$-$4 shown in Sect. \ref{SiOobs}, which is why we refrain from showing the channel maps here. We fitted 2D Gaussian components to each patch of emission above 20$\sigma_{\rm rms}$ in each of the mid-configuration channel maps covering these lines, using the SAD\footnote{http://www.aips.nrao.edu/cgi-bin/ZXHLP2.PL?SAD} (search and destroy) task in the AIPS package \citep{VanMoorsel1996}. The algorithm  filters out extended emission but indicates localised regions dominated by conditions which are much more favourable to SiO emission than the rest of the immediate vicinity. We selected Gaussians which were within either 0.01'', or, if larger, the position errors of their counterparts in adjacent channels to ensure selection of the strongest signal, as isolated components could be fitting artefacts. The position error of a fitted Gaussian component is approximately 0.5*(synthesised beam)/(signal to noise ratio) \citep{Condon1997,Richards2011}. Thus, for even a weak 10$\sigma_{\rm rms}$ component the accuracy is better than 0.015''. We show the spatial distribution and velocity of the SiO peak positions of the v=0 emission of the mid-configuration in Fig. \ref{maser}.
	
\begin{figure}[]
		\centering
		\includegraphics[width=8.5cm]{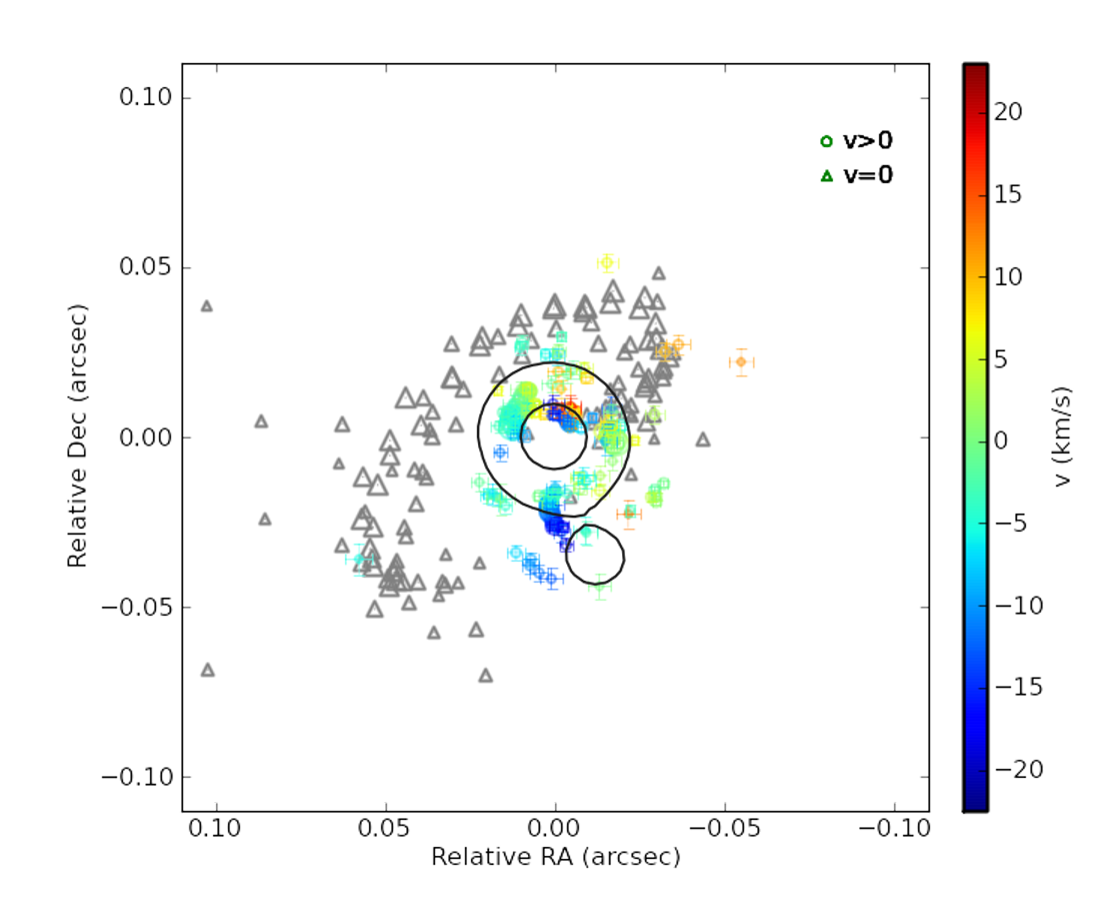}
		\caption{Same as Fig. \ref{maser}, but slightly zoomed-in and including the SiO v>0 lines. For the sake of readability, we have omitted the error bars on the v=0 data, and coloured them grey.}
		\label{maser2}
	\end{figure}

	The ground-state vibrational $^{\rm 28}$SiO $J=\ $5$-$4 v=1 and v=2, the $^{\rm 28}$SiO $J=\ $6$-$5 v=1 and v=3, the $^{\rm 29}$SiO $J=\ $6$-$5 v=1 and the $^{\rm 30}$SiO $J=\ $6$-$5 v=1 lines are also within our bands and we performed a similar analysis using the extended configuration images. These lines are generally strong, and more prone to masing. For the more sparse visibility-plane coverage, the position error of a fitted Gaussian component is (synthesised beam)/(signal to noise ratio), so even for a weak 10$\sigma_{\rm rms}$ component the accuracy is better than 0.005''. The results of the v>0 analysis are shown in Fig. \ref{maser2}.
	
	The SiO peak positions of the v=0 emission (Fig. \ref{maser}) outline an elongated, tilted, hook-like feature, with a thickness of $\sim$0.018'' and only a few outliers. It exhibits an opening around the region of the secondary continuum peak, and the highest speeds can be found at the end-points of the curve, with magnitudes up to 20$\kms$. The most red-shifted signal is found quite close to the west of the primary continuum peak, while the most blue-shifted  signal is found $\sim$0.03'' to the east of the secondary continuum peak. The v>0 results (Fig. \ref{maser2}) have velocities that are mainly concentrated around the systemic velocity, and encompass approximately the entire circumference of the primary continuum peak. However, a region of particular interest is identified to the south of the primary continuum peak. Here, a curved trail of entirely blue-shifted v>0 patches is found to connect the locations of the primary and the secondary continuum components. In addition, a shorter second tail is found to the east of the secondary. Combined, the v=0 and v>0 features appear consistent with the expected flow pattern of the wind material as it is subjected to the conditions caused by interaction with a companion. We elaborate on these ideas in Sect. \ref{rotdiscus}.
	
	\section{Discussion} \label{discus}
	
	\subsection{The nature of the spiral} \label{spiraldiscus}
	
	Using the spiral-arm spacing derived from the inner 5'' of the CO emission (Sect. \ref{COobs}), we can derive the inclination of the system by fitting the pattern in the CPmom0 with an inclined Archimedean spiral. The radial coordinate $r'$ of the spiral is given by
	\begin{equation}
	r' = R_{tilt}R_{incl}r,
	\end{equation}
	with $R_{tilt}$ and $R_{incl}$ the rotation matrices specifying the tilt of 6$^\circ$ (the PA of the system, see Fig. \ref{COstereo}) and the inclination of the system, respectively. In the initial coordinate system ($x$,$y$) define the plane of the sky, and $z$ the line of sight. And in this coordinate system $r$, the radial distance of the spiral arm from the continuum peak, is defined as
	\begin{equation}
	r = b\phi+a,
	\end{equation}
	with $b$ the spacing between the spiral arms, $\phi$ is the angle in the plane of the sky, and $a$ the phase of the spiral, assumed to start at the south of the primary continuum peak. To assess the quality of the fit, the line emission values (averaged over one beam width) in the data pixels at radii below 5'' crossed by the Archimedean spiral are averaged, and compared to similar averages for different inclination angles. The angle that yields the highest pixel average is assumed to be the inclination of the system, which was determined to be (38$\pm$3)$^\circ$. This value is in agreement with the previously visually constrained value of $\sim$40$^\circ$ by \citet{Doan2017}. 
	
	We confined this procedure to the inner 5'' because the data show that the distance between the windings increases as a function of distance from the star. Indeed, the spiral arms beyond 5'' are separated by up to 5'', the spiral arms within the 5'' radius region show a $\sim$2'' spacing (see Sect. \ref{COobs}), and the first few spiral windings can be measured to be only separated by a distance of $\sim$0.25'' (see Sect. \ref{HCNobs}). This is not to be expected from the typical mathematically ideal Archimedean spiral with which such data are typically compared. 
	
	Hydrodynamical models show that when the mass-losing star and the companion have comparable masses, the resulting spiral morphology does not follow an ideal Archimedean trend, and the distance between the spiral arms will vary as a function of distance from the AGB star. This is because the spiral-forming mechanism in such a fully detached system, implying no (wind-)Roche-lobe overflow, consists of a multitude of interfering phenomena, which can be properly discerned in, for example, the models of \citet{Liu2017}.
	
	The wind is beamed in the immediate vicinity of the companion by its gravitational potential, resulting in a density- and velocity-enhanced Bondi-Hoyle-Lyttleton (BHL) tail \citep{Hoyle1939,Bondi1944,Bondi1952}. This tail trails behind the companion, and produces a spiral shape with an orientation opposite to the motion of the companion. That is to say, if the spiral tail opens-up in a clockwise fashion (such as for $\pi^1$ Gruis), then this implies a counter-clockwise orbital motion of the companion.
	The edge of this tail oriented in the direction of the companion's motion has a high speed, and shocks the wind material in front of it, leaving an overdense shocked region in its wake. 
	The mass-losing star is also orbiting around the system's centre of mass. 
	This wind collides with the BHL tail, forming a low-velocity, high density spiral-shaped wake in the wind. 
	The faster spiral front will eventually catch up with the slower one, and sometimes merge into a single new front (see first panel of Fig. 5 in \citet{Liu2017}), or, if the shock is too strong, back into a slower and a faster front, which collide again at larger radii (see Fig. 3 in \citet{Liu2017}). Furthermore, the slowly accelerating winds of M- and S-type stars \citep[e.g.][]{Decin2006,Decin2018} would also steadily increase the distance between consecutive windings at further radii (see Sect. \ref{dynamics}). 
	
	
	\subsection{Anatomy of the inner wind} \label{rotdiscus}
	
	The gas around the primary continuum peak with a line of sight velocity close to the systemic velocity probed by the resolved SiO emission and the HCN emission (See Figs. \ref{HCNmom} and \ref{SiOmom}, respectively) traces a strip with a PA of $\sim$55 degrees within 0.1'' of the star. Beyond this distance, the PA changes to $\sim$96 degrees and remains unchanged all the way to the largest spatial features traced by CO.
		
	When viewing a naked inclined EDE with a tangential velocity field, the gas close to the systemic velocity will lie along the minor axis
	of the projected view of the EDE. This stands in contrast to a radial velocity field, for which the gas around the systemic velocity will lie along the major axis of the projected EDE. Hence, a smaller-scale tangential field embedded in a larger-scale radial field (i.e. expanding medium) will result in a combined effect, whose properties depend on the relative strength and size of both fields.	This means that the PA of the $\sim$zero Doppler-velocity gas in the inner wind traced by SiO and HCN can be explained by a gradual shift in velocity regime, from the larger-scale radial field to an inner velocity field that is more rotational in nature.
	
	HCN and SiO probe the gas on length scales of the order of the binary separation. This means that they not only probe the post-interaction gas that is freely flowing outward (such as seen in CO), but also the pre- and peri-interaction gas, that is undergoing strong hydrodynamical perturbations due to the inferred binary interaction. Hydrodynamical models of spiral-inducing AGB binaries that focus on the wind-companion interaction zone \citep{Mastrodemos1998,Mohamed2012,Saladino2019} show that, before transitioning to a radial wind, most stream lines within the orbital radius of the companion are nearly parallel to the orbital displacement vector (see e.g. Fig. 5 in \citet{Saladino2019}). Thus, the Doppler shifts in the line emission from this zone will not exclusively exhibit the typical features of a radial wind, but rather a mixed signal that includes signatures associated with rotation. Hence, as a first scenario, we propose that the detected feature could be consistent with what is expected from the spiral-forming region.

    There is, however, another scenario that can explain the kink of the zero Doppler-velocity region of the moment1 maps of SiO and HCN. It could be tracing an inclined, tilted, compact, and rotating EDE with a radius of a dozen or so stellar radii ($\sim$0.1''). Such discs have been previously detected around semi-regularly pulsating AGB stars \citep{Kervella2016,Homan2017,Homan2018}, so it would not be implausible to find one here. The results of the analysis of the SiO peaks and masers in Sect. \ref{maserdiscus} provides strong evidence against the above mentioned disc hypothesis. Assuming the v=0 hook (see Fig. \ref{maser}) is the signature of a compact and rotating disc, then it is difficult to explain the large gap to the south, as such a gap would have been filled-up by the disc's differential rotation in less than two years, assuming a maximal speed of $\sim$20~$\kms$. The hook-like feature more consistently resembles the first half-winding of the clockwise spiral. In addition, the highest speeds are expected to trace the entire inner rim of the disc, i.e. a radial gradient is expected to be seen. In this case, the highest speeds are found only at the end-points of the hook. This is not consistent with the dynamics of an inclined differentially rotating EDE, which will always exhibit its maximal velocities adjacent to the continuum peak, and along the major axis of the projected view of the disc. Lower velocities would be found at larger radii along the same axis, even when embedded in a larger-scale radial field \citep{Homan2016}. Finally, the orientation of this disc would imply a projected angular momentum vector oriented along the north-east to south-west axis, and thus the orbital plane of the system perpendicular to it. This orientation is very difficult to reconcile with that of the symmetry axes of the larger-scale nebula (see e.g. Fig. \ref{COstereo}).
    Combined, these arguments thus favour the aforementioned spiral hypothesis. 
    
    \subsection{Maser emission and mass of tentative companion} \label{maserdiscus}
    
    The map of the v>0 maser emission reveals that most SiO masers are located in a ring around the star, at radii less than the distance to the secondary, and with projected speeds close to the systemic velocity. This is expected for an accelerating outflow which produces tangential beaming \citep{Chapman1985}. Within the inner few stellar radii, i.e. before dust has formed, material producing SiO masers may be accelerated by shocks from stellar pulsations. If the material has zero velocity at the stellar surface, and reaches typically 7~$\kms$ at 3-5 stellar radii from the star, this implies a sporadic outward acceleration $\sim$1~$\kms$ per au. The material then falls back under the star's gravity, so that the net outflow across an SiO maser zone is only $\sim$0.5~$\kms$ \citep{Assaf2018} in the absence of binary interaction. However, to the south of the primary continuum peak, an anomaly manifests itself as a row of progressively more blue-shifted masers that connect the primary to the secondary continuum peak. SiO masers are known to sometimes exhibit such features, caused by the interaction between pulsation, dust formation and exponential amplification of small differences in conditions \citep[e.g.][]{Gray2009,Assaf2013}. These are usually seen at various position angles and their velocity may increase or decrease with distance from the star.
    	
    In order to generate further insight for future investigation we tested the hypothesis that this feature may be tracing the flow of gas from the AGB star to a companion. This type of mass transfer has been previously observed around in the R Aqr system \citep{Bujarrabal2018}, where gas is seen to be flowing from the AGB star to a white dwarf companion, located at a distance of $12\pm2$~au.. The large-scale orientation of the $\pi^1$Gru nebula with the southern portion oriented towards Earth, would imply the same for the tentative companion, and explain the blue-shift. In addition, the material further from the star is more blue-shifted from --5~$\kms$ at a angular separation of $\sim$0.014'' to --20~$\kms$ at an angular separation of $\sim$0.034'' from the primary continuum peak. This gradual increase in projected speed could be due to simple gravitational acceleration towards the companion. We can use the measured acceleration of the gas as a way to estimate the mass of the companion. The projected distance of the feature is $\sim$2.7~AU, so assuming an inclination angle of 38$^\circ$ the total length of the feature is approximately 4.4~AU. Basic kinematics thus gives a mean acceleration $a$ of $\sim$4.5$\times$10$^{\rm -4}$~$ms^{-2}$. Assuming the gas is inside the companion's potential well and undergoing free fall, we equate this acceleration to the theoretically expected mean gravitational acceleration over this distance, i.e.
    \begin{equation}
     a = \frac{GM}{d_2-d_1} \int_{d_1}^{d_2} \frac{dr}{r^2},
    \end{equation}
    with $G$ the universal gravitational constant, $M$ the mass of the companion, and $d_1$ and $d_2$ the distances from the secondary continuum peak to where the acceleration extrema are measured. This results in a derived companion mass of $\sim$2.2$\times$10$^{\rm 30}$~kg, or approximately 1.1~$\mso$, consistent with the rough estimate made by \citet{Mayer2014}. This can be considered a rough upper-limit, since no additional effects that contribute to the acceleration of the gas (e.g. pulsations, dust, ...) have been taken into account.
    
    The v=0 loop (see Fig. \ref{maser}) does not appear to start (blue-shifted signal) at the location of the tentative companion, and appears to shear very close to the AGB star's surface, especially at higher red-shifts. Under the companion-induced spiral scenario, this can be explained by a combination of projection effects and extreme physics. The system's inclination angle of $\sim$40$^\circ$ compresses the projected spiral in the vertical direction, giving the impression that it is shearing the AGB star, while it is probably situated an appreciable distance behind the star. In addition, the thermodynamical conditions nearby a $\sim$1~$\mso$ companion (see Sect. \ref{maserdiscus}) in an AGB wind can be rather extreme \citep[e.g.][]{Saladino2019}. Therefore, the absence of v=0 SiO signal between the location of the tentative companion and the most blue-shifted data points is probably indicative of the absence of suitable conditions for the generation of such emission, rather than the absence of matter.
    
\begin{figure}[htp!]
    	\centering
    	\includegraphics[width=7.0cm]{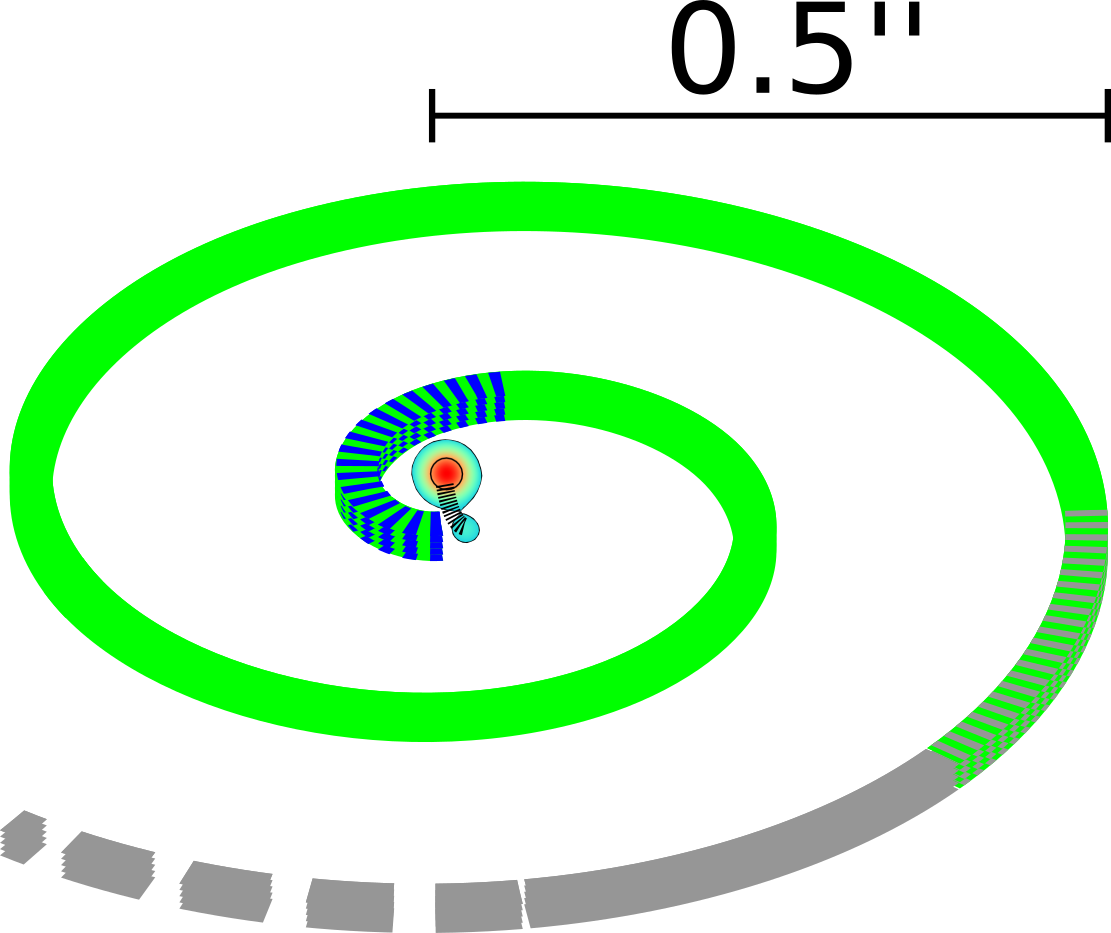}
    	\caption{Scematic diagram of the inner wind of $\pi^1$Gru. In the center is a cut-out of the 120$\sigma_{\rm rms}$ contour of the ALMA continuum, from which the inclined spiral is launched. The colours of the spiral represent the different diagnostics used: blue is the SiO, green is the HCN, grey is the CO emission, and black is the stream of accelerating gas probed by the maser emission. The dashes parts represent regions of overlap.}
    	\label{diagram}
    \end{figure}

    
    To better visualise the inner structure of the EDE, we present a diagram containing the principal features of the inner wind of the $\pi^1$Gru nebula in Fig. \ref{diagram}.
    
	\subsection{Radial wind dynamics} \label{dynamics}
	
	Accounting for the system's inclination of $\sim$38$^\circ$ (see Sect. \ref{spiraldiscus}), the mean speed of the northern arc of the HCN spiral is $\sim$12~$\kms$, at a deprojected distance of $\sim$90~au. This is substantially lower than the velocity extent of the CO, which is difficult to unambiguously deduce since the EDE signal and the hourglass signal are merged in the 15 to 30~$\kms$ range of projected velocities, or $\sim$20 to 40~$\kms$ actual speed. The substantial difference between the gas speed probed by the HCN and the CO emission implies that the gas is still accelerating in the region between 0.5'' and 10''. 
	
	We can use such dynamical constraints provided by the resolved HCN and CO maps to evaluate the exponent of the velocity beta-law approximation for a radial wind dynamics profile, given by
	\begin{equation}
	v(r)=v_\infty\left(1-\frac{r_{\rm dc}}{r}\right)^\beta,
	\end{equation}
	where $v_\infty$ is the terminal wind speed, $r_{\rm dc}$ is the dust condensation radius of the system, and $\beta$ is the parameter that determines the rate of acceleration. To this end we combine the above measurement with mean deprojected speeds and distances of the southern HCN arc, as well as the most prominent arcs in the red-shifted wing of the CO cube. The measurements are summarised in Table \ref{arcs}. The mean distance measurement is made at the arcs' mean speed. The uncertainty on the distance is measured as the 3$\times \sigma_{\rm rms}$ width of the arm at the mean speed. The uncertainty on the mean speed is determined by the velocity extent over which the arc signal persists. The latter is certainly an overestimate: the large equatorial thickness angle of the EDE and the considerable inclination angle of the system conspire to give the arc signals a large velocity width. However, because we cannot disentangle the velocity-extent of the signal from the geometry of the EDE without a velocity model, we keep this uncertainty as upper limit.
	
\begin{figure}[]
        \centering
        \includegraphics[width=8.5cm]{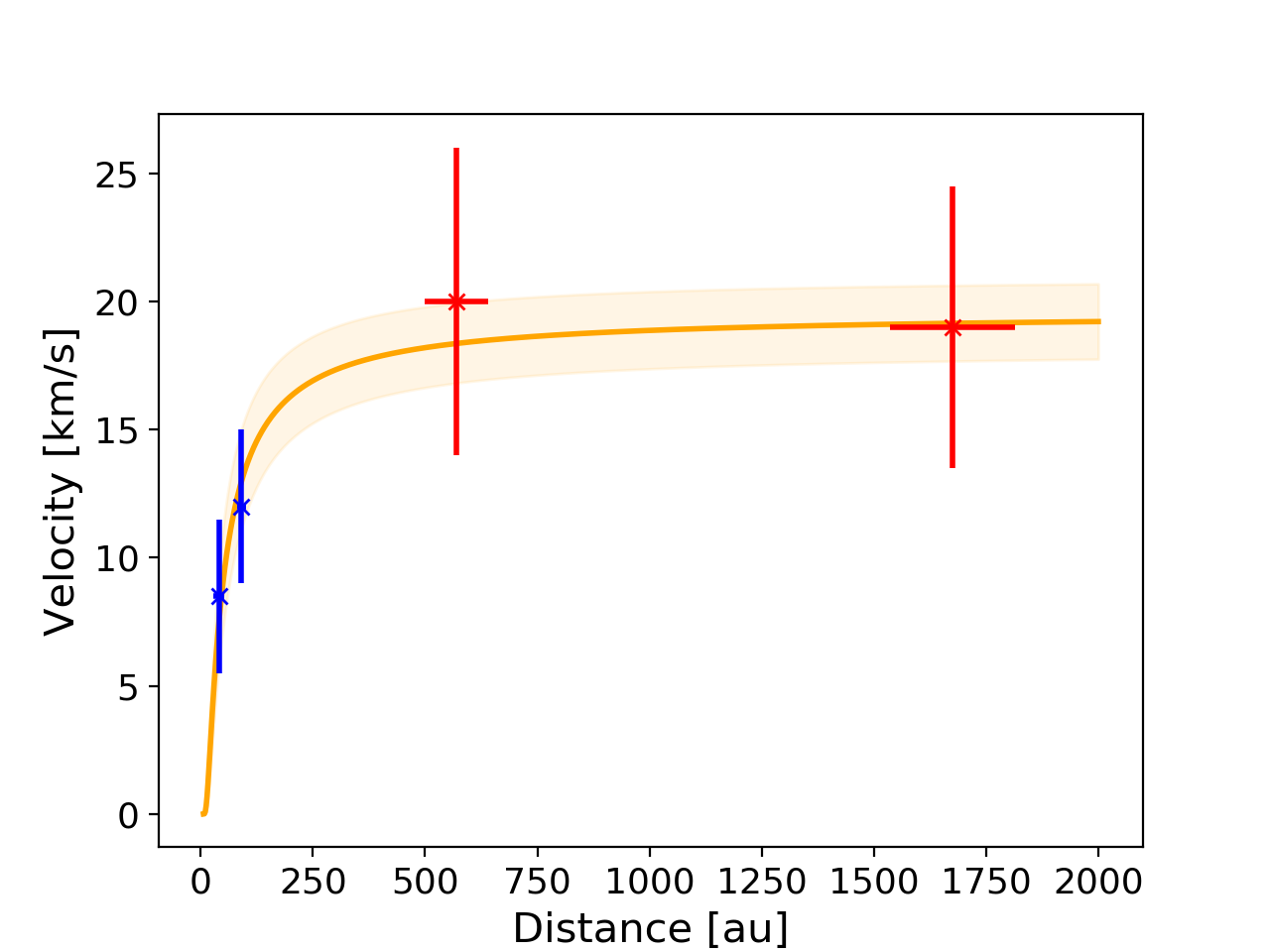}
        \caption{Measurements of the dynamics in the EDE. The blue (red) markers represent the most prominent arcs of the HCN (CO) emission. The orange curve represents the best-fit beta-law. The lighter coloured zone around the curve represents the 1$\sigma$ uncertainty on the derived parameters}
        \label{betalaw}
    \end{figure}

	Assuming that the dust condensation radius is around 6~au (see Montarg\`{e}s et al., \emph{in prep.}), we used the SCIPY.ODR python library to derive a beta-law best-fit with a terminal velocity value of $\sim$19.5$\pm$1.4~$\kms$, and an exponent of $\beta$=6.0$\pm$1.1, as shown in Fig. \ref{betalaw}. Such high values for $\beta$ have been found for some M-type AGB stars \citep[e.g.][Gottlieb et al. \emph{in prep.}]{Decin2006,Khouri2014,Decin2018}, and are not well understood. Many scenarios have been proposed to explain such a slow acceleration. For example, the inefficient coupling of the stellar radiation field to the oxygen-rich dust in the CSE \citep{Hofner2008,Bladh2019}, a radius-dependent dust formation efficiency, which triggers the formation of different dust types at different radii \citep{ElMellah2020}, the formation of large quantities of dust grains with a low fractal dimension \citep{Decin2018}, etc.
	
	Using these dynamical constraints, we can deduce the orbital properties of the binary system that are responsible for the observed spiral pattern. Accounting for the acceleration profile derived above, the spiral windings closest to the AGB star have a dynamical crossing time scale of $\sim$15 years. Using Kepler's third Law of planetary motion, and assuming a 1.5~$\mso$ system mass \citep{Siess2006,Siess2008}, we deduce an orbital separation between the AGB star and the companion of $\sim$7~au. Under projection due to the system's inclination, this object would appear at a distance from the AGB star of $\sim$6~au, or $\sim$0.04'' along the north-to-south axis. This coincides precisely with the position of the secondary component observed in the ALMA continuum (see Fig. \ref{cont}), strengthening the case that the secondary continuum peak is directly related to the presence of a companion.
	
	When applying the same argument to the spiral arcs observed at larger radii, which have a typical separation of $\sim$2.5'', the dynamical crossing time scales augment to $\sim$100 years. This is a clear indication the geometrical nature of the spiral changes as it moves outward, probably due to the complex hydrodynamical interplay of the phenomena described in Sect. \ref{spiraldiscus}. 
	
	\begin{table}
		\caption{Deprojected speed-distance pair measurements for different spiral arcs in order to determine the radial dynamics of the wind along the equatorial plane.}
		\centering          
		\label{arcs}
		\begin{tabular}{llll}
			\hline\hline
			\noalign{\smallskip}
			Molecule & Arc      & Mean speed   & Mean dist.  \\
			         & position & [$\kms$]    & to AGB [au]  \\
			\noalign{\smallskip}
			\hline    
			\noalign{\smallskip}
			HCN & S               & 8.5  $\pm$ 3.0 & 41 $\pm$ 12 \\
			\noalign{\smallskip}
			HCN & N               & 12.0 $\pm$ 3.0 & 90 $\pm$ 10 \\
			\noalign{\smallskip}
			CO  & N, last but one & 20.0 $\pm$ 6.0 & 570 $\pm$ 70 \\
			\noalign{\smallskip}
			CO  & N, farthest     & 19.0 $\pm$ 5.5 & 1675 $\pm$ 140 \\
			\hline    
		\end{tabular}
		\vspace{1ex}\\
		\begin{flushleft}
			\textbf{Note}: S stands for south, N for north.
		\end{flushleft}
	\end{table}

	\section{Summary} \label{summ}
	
	The wind of the S-type AGB star $\pi^1$Gruis was observed with ALMA as part of the {\sc Atomium} large programme. The combination of three configurations ensured a continuous covering of spatial scales from $\sim$0.01'' to $\sim$9.3'', permitting an in-depth morphological investigation of the star's circumstellar nebula. To this end we presented the continuum emission, and the CO v=0 $J=\ $2$-$1, HCN v=0 $J=\ $3$-$2, and the SiO v=0 $J=\ $5$-$4 molecular line emission. The continuum shows two maxima, distinctly separated at the 120$\times \sigma_{\rm rms}$ level. The fainter secondary component is located about 0.04'' to the south-west of the brightness maximum, which corresponds to a physical distance of $\sim$6~au. Analysis of the CO emission confirms the previously determined morphological character of the nebula: the bulk of the gas is confined into an inclined, radially outflowing equatorial density enhancement harbouring a spiral. We analysed the geometrical properties of the spiral and found that the emission at radii below 5'' can best be explained by a non-eccentric, clockwise Archimedean spiral, inclined under an angle of $\sim$()38$\pm$3)$^\circ$ with respect to the line-of-sight. Analysis of the HCN emission, which probes the wind at scales below $\sim$1.5'', unambiguously confirms this, as it can be seen to trace the first two spiral windings. A dynamical analysis of the speed of the spiral arm at different radii permitted us to deduce a terminal velocity of $\sim$19.5$\pm$1.4~$\kms$ and an acceleration parameter $\beta$ of $\sim$6.0$\pm$1.1. The dynamical crossing time-scale of the spiral arms in the inner wind is consistent with a companion located at a projected distance of 0.04'' to the south of the AGB star, precisely where the secondary component in the continuum is located. Further analysis of the SiO data reveals that at length scales below 0.3'', the Doppler-shifts of the emission are consistent with gas in rotation. Using a method to filter-out super-compact SiO emission, we used different vibrational states of the $^{\rm 28}$SiO $J=\ $5$-$4, $^{\rm 29}$SiO $J=\ $5$-$4, $^{\rm 28}$SiO $J=\ $6$-$5, and $^{\rm 30}$SiO $J=\ $6$-$5 lines to map the gas dynamics within a radius of 0.05''. This revealed a hook-like feature starting approximately at the location o f the secondary continuum peak, and curling northwards in a clockwise fashion. This is consistent with companion-induced spiral tail found in hydrodynamical simulations in the literature. The v>0 SiO maser emission exposed what could be interpreted as a stream of gas flowing from the AGB star to the potential companion, located at the secondary continuum brightness peak. By measuring the acceleration of the gas, we derived a tentative upper limit on the mass of the companion of 1.1~$\mso$.
	
	
	\begin{acknowledgements}  
		
		We would like to sincerely thank the Referee, whose constructive suggestions have substantially improved the paper.
		
		This paper makes use of uses the following ALMA data: ADS/JAO.ALMA\#2018.1.00659.L, ‘ATOMIUM: ALMA tracing the origins of molecules forming dust in oxygen-rich M-type stars’. ALMA is a partnership of ESO (representing its member states), NSF (USA), and NINS (Japan), together with NRC (Canada),  NSC and ASIAA (Taiwan), and KASI (Republic of Korea), in cooperation with the Republic of Chile. The Joint ALMA Observatory is operated by ESO, AUI/NR.A.O, and NAOJ.
		
		W.H., S.H.J.W., L.D, M.M. acknowledge support from the ERC consolidator grant 646758 AEROSOL. 
		
		This project has received funding from the European Union's Horizon 2020 research and innovation programme under the Marie Sk\l{}odowska-Curie Grant agreement No. 665501 with the research Foundation Flanders (FWO) ([PEGASUS]$^2$ Marie Curie fellowship 12U2717N awarded to M.M.). 
		
		TD acknowledges support from the Research Foundation Flanders (FWO) through grant 12N9920N. 
		
		TJM is grateful to the STFC for support via grant number ST/P000312/1.
		
		JMCP acknowledges funding from the UK STFC (grant number ST/T000287/1) 
		
		MVdS acknowledges support from the Research Foundation Flanders (FWO) through grant 12X6419N.
		
		S.E. acknowledges funding from the UK Science and Technology Facilities Council (STFC) as part of the consolidated grant ST/P000649/1 to the Jodrell Bank Centre for Astrophysics at the	University of Manchester.
		
		We acknowledge financial support from “Programme National de Physique Stellaire” (PNPS) of CNRS/INSU, France.
		
		We used the SIMBAD and VIZIER databases at the CDS, Strasbourg (France)\footnote{Available at \url{http://cdsweb.u-strasbg.fr/}}, and NASA's Astrophysics Data System Bibliographic Services. This research made use of IPython \citep{PER-GRA:2007}, Numpy \citep{5725236}, Matplotlib \citep{Hunter:2007}, SciPy \citep{2020SciPy-NMeth}, and Astropy\footnote{Available at \url{http://www.astropy.org/}}, a community-developed core Python package for Astronomy \citep{2013A&A...558A..33A}.

	\end{acknowledgements}

	\bibliographystyle{aa}
	\bibliography{wardhoman_biblio}

	\IfFileExists{wardhoman_biblio.bbl}{}
	{\typeout{}
		\typeout{******************************************}
		\typeout{** Please run "bibtex \jobname" to obtain}
		\typeout{** the bibliography and then re-run LaTeX}
		\typeout{** twice to fix the references!}
		\typeout{******************************************}
		\typeout{}
	}
	
	\clearpage
	\begin{appendix}
		
		\section{Global morphological properties of the nebula from CO emission} \label{A.CO}
		
\begin{sidewaysfigure*}[]
			\centering
			\includegraphics[width=12cm]{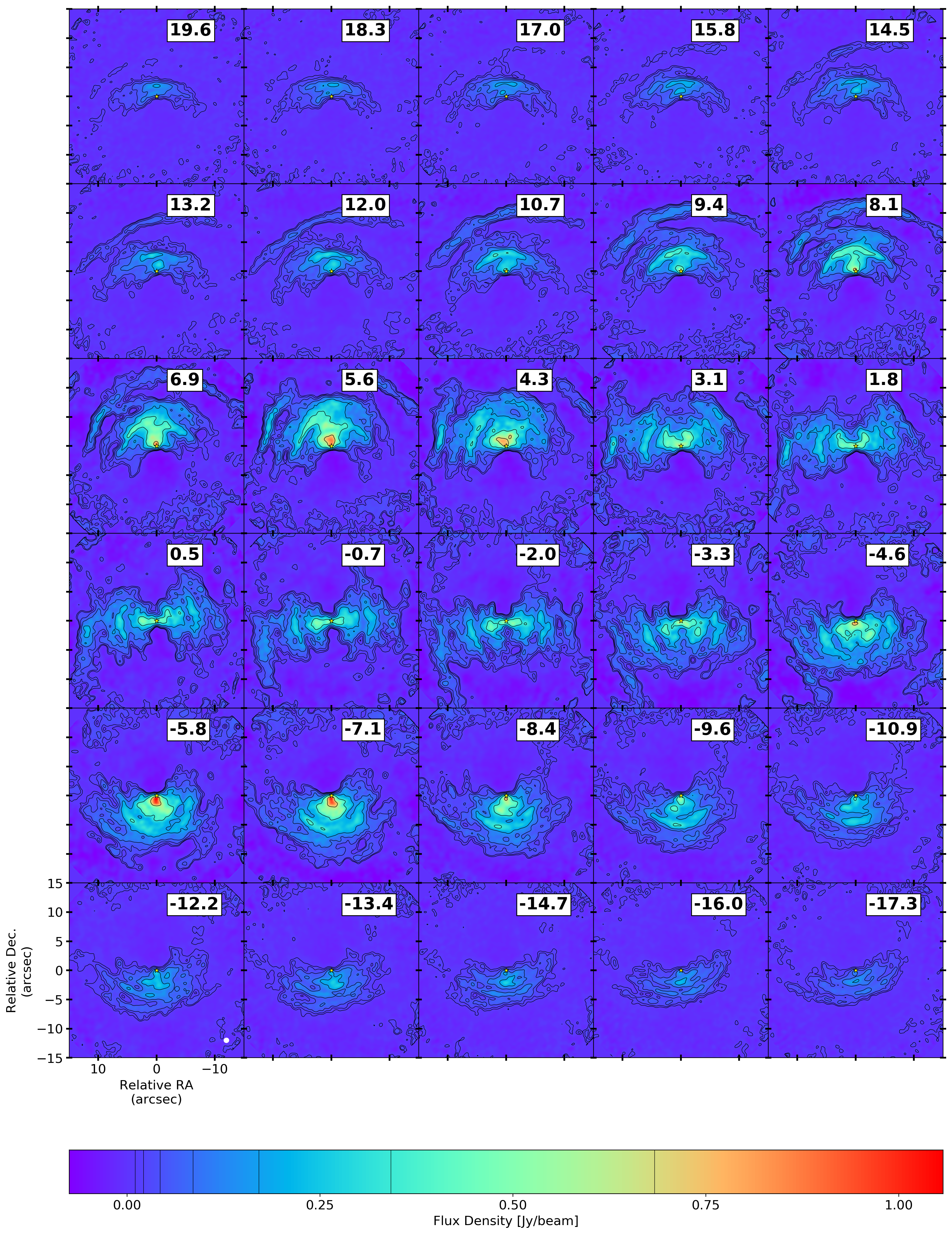}
			\includegraphics[width=12cm]{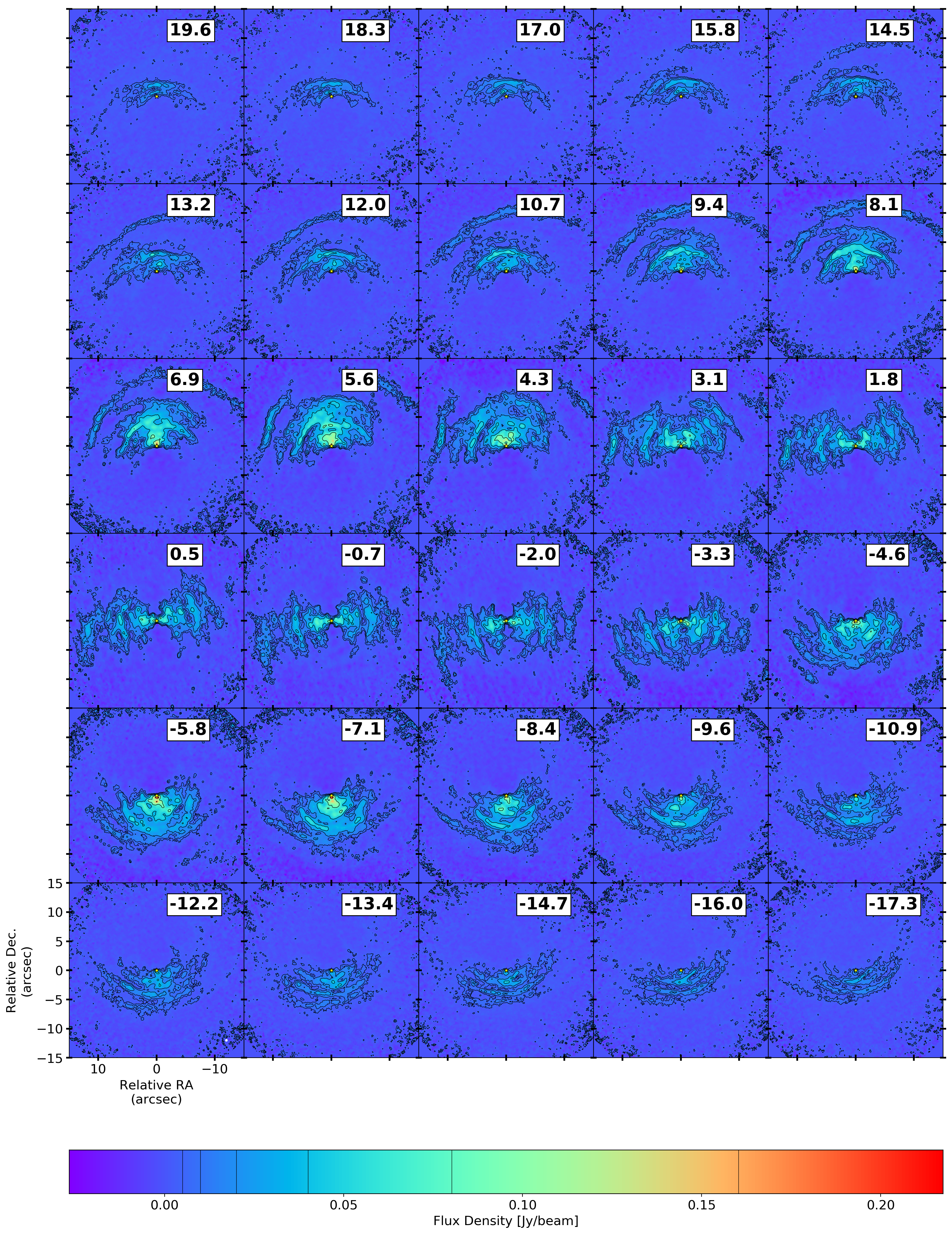}
			\caption{Channel maps showing the resolved emission of the CO v=0 $J=\ $2$-$1 line in the $\pm \sim$18~$\kms$ velocity range. The labelled velocities have been corrected for $v_{*}$=-12~$\kms$. Contours are drawn at 3, 6, 12, 24, 48, 96, and 192$\times \sigma_{\rm rms}$. Length scales and ALMA beam are indicated in the bottom left panel. The maps are centred on the continuum peak position, which is indicated by the yellow star symbol. \emph{Left:} Compact 12m array data. $\sigma_{\rm rms}$ (= 3.6$\times {\rm 10}^{\rm -3}$ Jy/beam), beam size = 0.807''$\times$0.768''. \emph{Right:} Combined 12m array data. $\sigma_{\rm rms}$ (= 1.7$\times {\rm 10}^{\rm -3}$ Jy/beam), beam size =  (0.292''$\times$0.278'')}
			\label{COchan}
		\end{sidewaysfigure*}

\begin{figure}[]
			\centering
			\includegraphics[width=8cm]{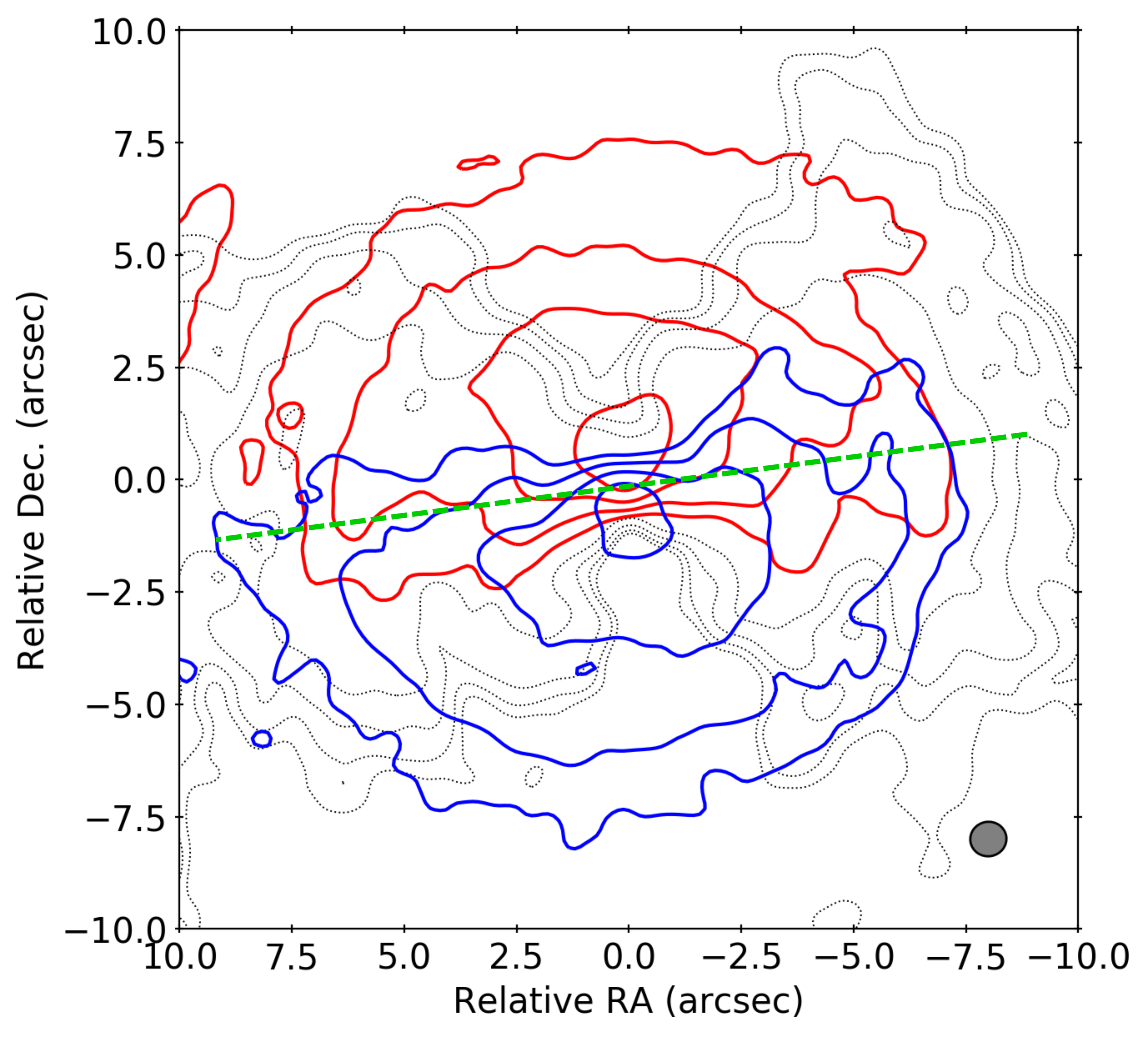}
			\caption{Stereogram of the cube shown in Fig. \ref{COchan}. Contours are drawn every 3, 6, 12, and 24 times the rms noise value in the spectral region of the bandpass without detectable line emission ($\sigma_{\rm rms}$ = 3.6$\times {\rm 10}^{\rm -3}$ Jy/beam). The channel at stellar velocity is shown in black dotted contours, and the mean red (blue)-shifted emission in red (blue). The green dashed line represents the symmetry axis of the map. The ALMA beam is shown in the bottom right (0.807'' $\times$ 0.768'').}
			\label{COstereo}
		\end{figure}

		The stereogram of the velocity domain in Fig. \ref{COchan} is shown in Fig. \ref{COstereo}. The black-dashed contours trace the five channels around stellar velocity, whereas the red and blue contours trace the velocity-integrated emission of the respective Doppler-shifted parts of the data. It shows that the emission around stellar velocity has a distinct bowtie-shape. 
		This effectively represents a cross-section through the EDE, as does the zero Doppler-velocity channel in Fig. \ref{COchan}. The position-angle of the system can be readily deduced from the orientation of the strip of gas with projected speeds close to the systemic velocity, or from the relative positions of the red and blue contours: It is tilted counter-clockwise around the continuum peak over a visually constrained angle of about 6$^\circ$, i.e. the symmetry axis of the stereogram and moment1 maps has a position-angle of 96$^\circ$ (measured north-to-east).
		
\begin{figure*}[]
			\centering
			\includegraphics[width=8cm]{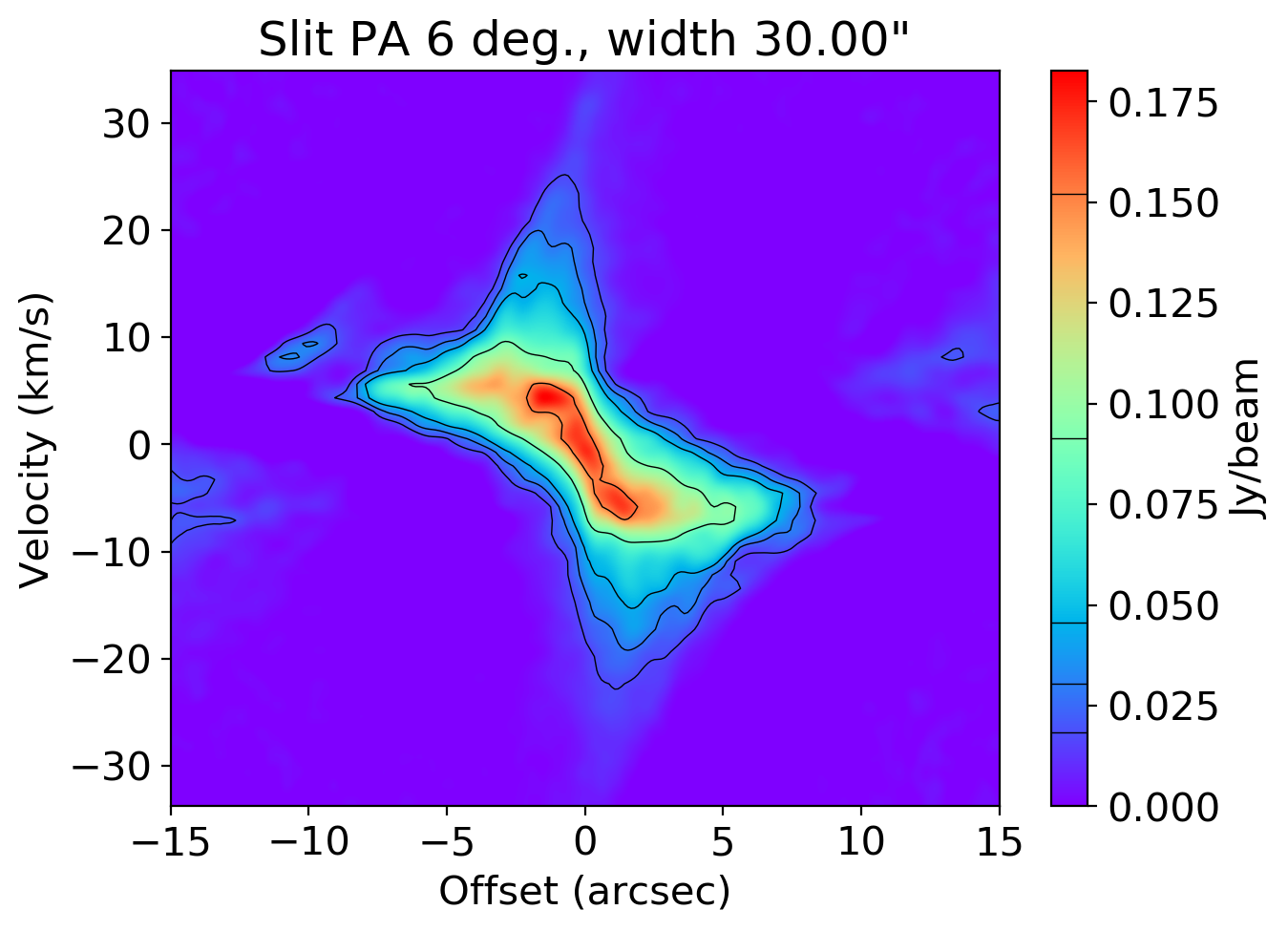}
			\includegraphics[width=8cm]{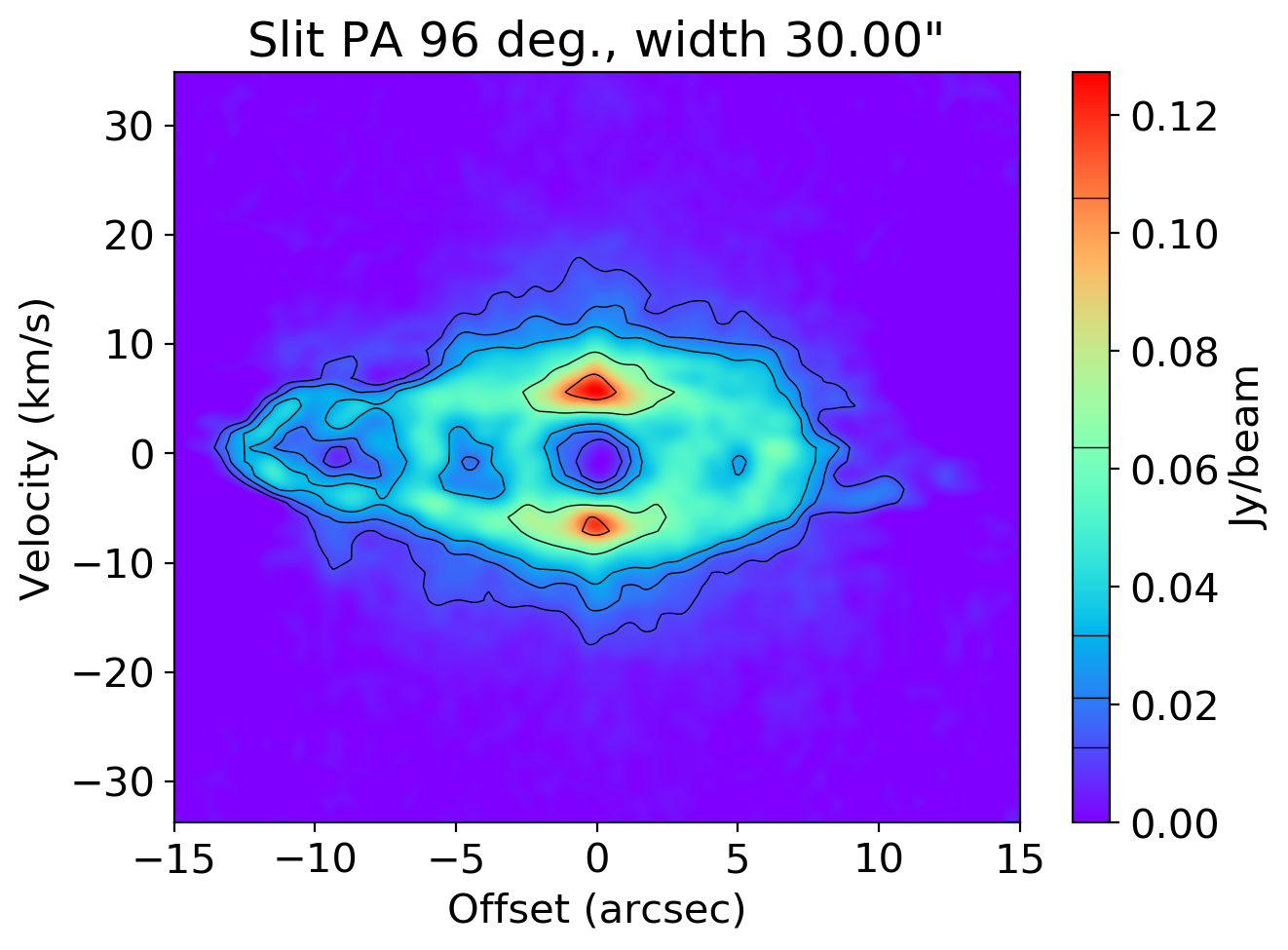}
			\caption{Orthogonal wide-slit position-velocity diagrams of the entire CO v=0 $J=\ $2$-$1 cube. The left panel is constructed with a slit along the PA = 6$^\circ$ axis, and the right panel with a slit orthogonal to that used in the left panel.}
			\label{COpvd}
		\end{figure*}

		Fig. \ref{COpvd} shows the "wide-slit" position-velocity diagrams of the data, with a slit width equal to the total width of the field of view. The diagram constructed by taking the slit along a PA of 6$^\circ$ exhibits the typical pattern associated with EDEs, independent of their intrinsic velocity distribution \citep{Homan2016}. Its orthogonal counterpart reveals a pattern with two bright spots around $\pm {\rm 10}\ \kms$, tracing the edges of a distinct circular zone of much lower emission. This latter signature is unambiguously associated with a radial velocity-field, and is distinctly different from a rotating field \citep{Homan2016}.
		
\begin{figure}[htp]
			\centering
			\includegraphics[width=8.5cm]{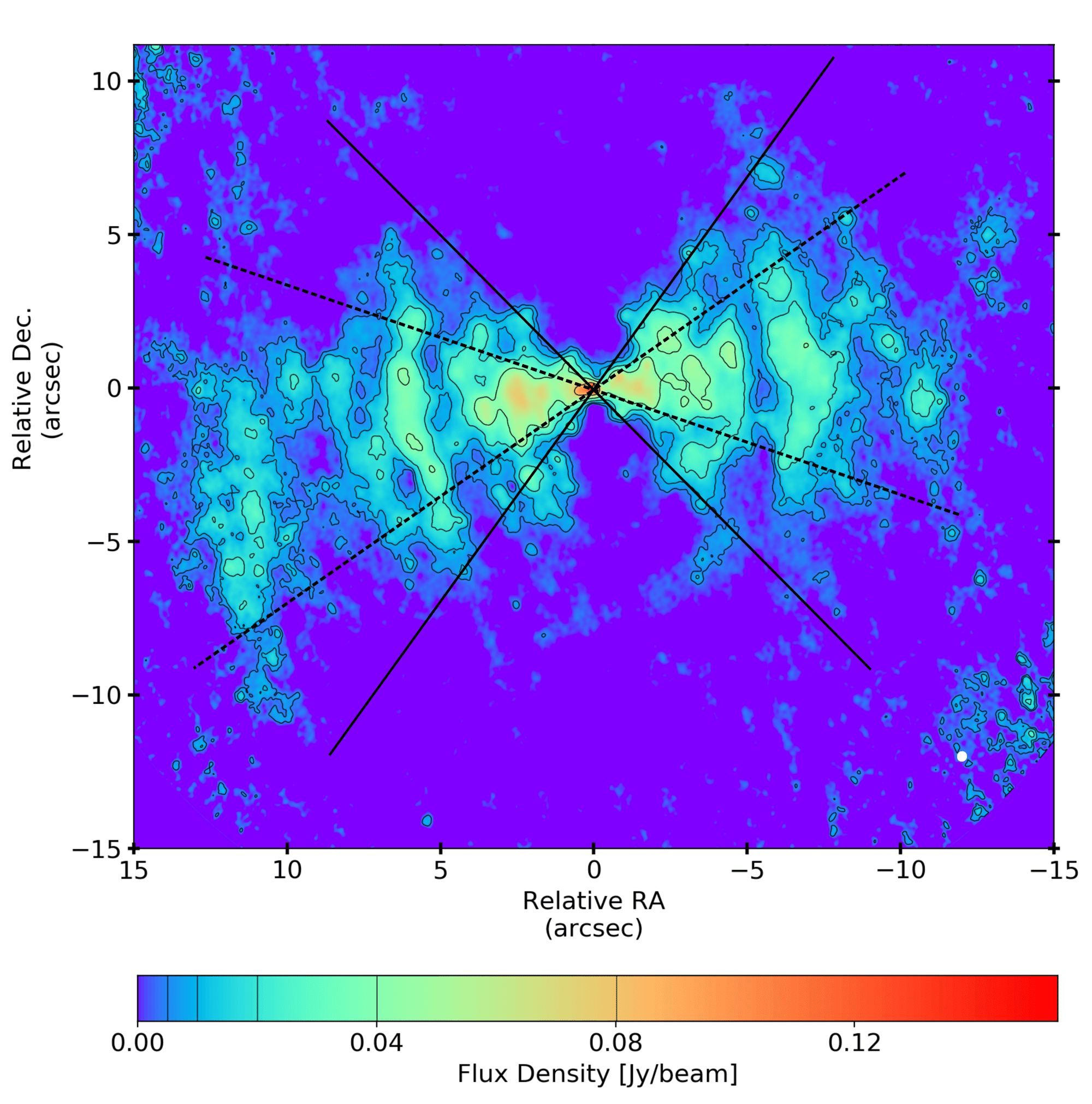}
			\caption{Central channel of the combined CO dataset, showing how the equatorial thickness angles have been measured. The area bound by the full(dashed) black lines is measured w.r.t the 3(12)$\times \sigma_{\rm rms}$ contours. It is measured as the angle bound by the lines crossing the equatorial plane.}
			\label{COopening}
		\end{figure}

\begin{figure}[htp]
			\centering
			\includegraphics[width=8.5cm]{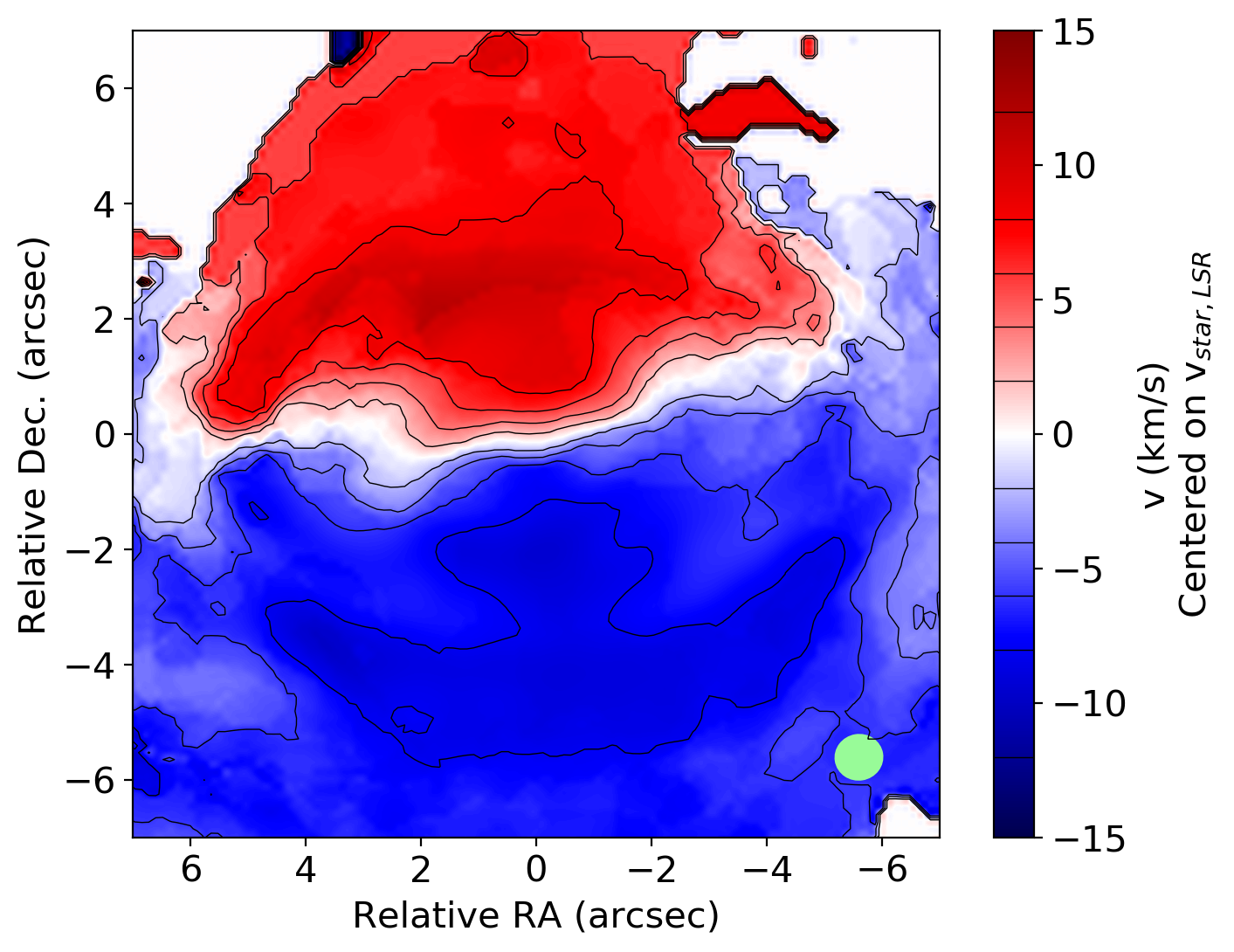}
			\caption{moment1 map of the CO emission in the $\pm$20~$\kms$ velocity range. Only emission with a signal above 9$\times \sigma_{\rm rms}$ is shown.}
			\label{COmom1}
		\end{figure}

		Using the emission in the central channel it is possible to quantify the thickness of the EDE with respect to the equatorial plane as an opening angle w.r.t. the symmetry axis of the stereogram or moment1 maps (see Figs. \ref{COstereo} and \ref{COmom1}), which shall henceforth be referred to as the equatorial thickness angle. The 3$\times \sigma_{\rm rms}$ contour constrains this angle to $\sim$100$^\circ$ (or $\pm$50$^\circ$ from the orbital plane, see Fig. \ref{COopening}) from the emission within a distance of 5'' from the AGB star. 
		The 3$\times \sigma_{\rm rms}$ contour includes a large portion of the weaker, more nebulous and erratic emission. Considering the highly complex hydrodynamical instability features that are expected to arise in a medium where a dense EDE is sheared by a polar high-velocity component (see \citet{Doan2020}), the 3$\times \sigma_{\rm rms}$ contour may arguably represent material that is only loosely associated with the actual EDE. To minimise this issue, we also measure the equatorial thickness angle at the 12$\times \sigma_{\rm rms}$ level to approximately 55$^\circ$, 15$^\circ$ larger than the 40$^\circ$ \emph{opening angle} deduced by \citet{Doan2020}.
		
		Finally, for the sake of completeness, we also show the moment1 map (Fig. \ref{COmom1}) of the same velocity domain. The moment1 map also demonstrates the systematic offset between the red and blue-shifted portions of the emission.
		
	\section{Higher velocity hourglass signature in the line wings} \label{A.hg}
	
\begin{figure*}[]
		\centering
		\includegraphics[width=14.cm]{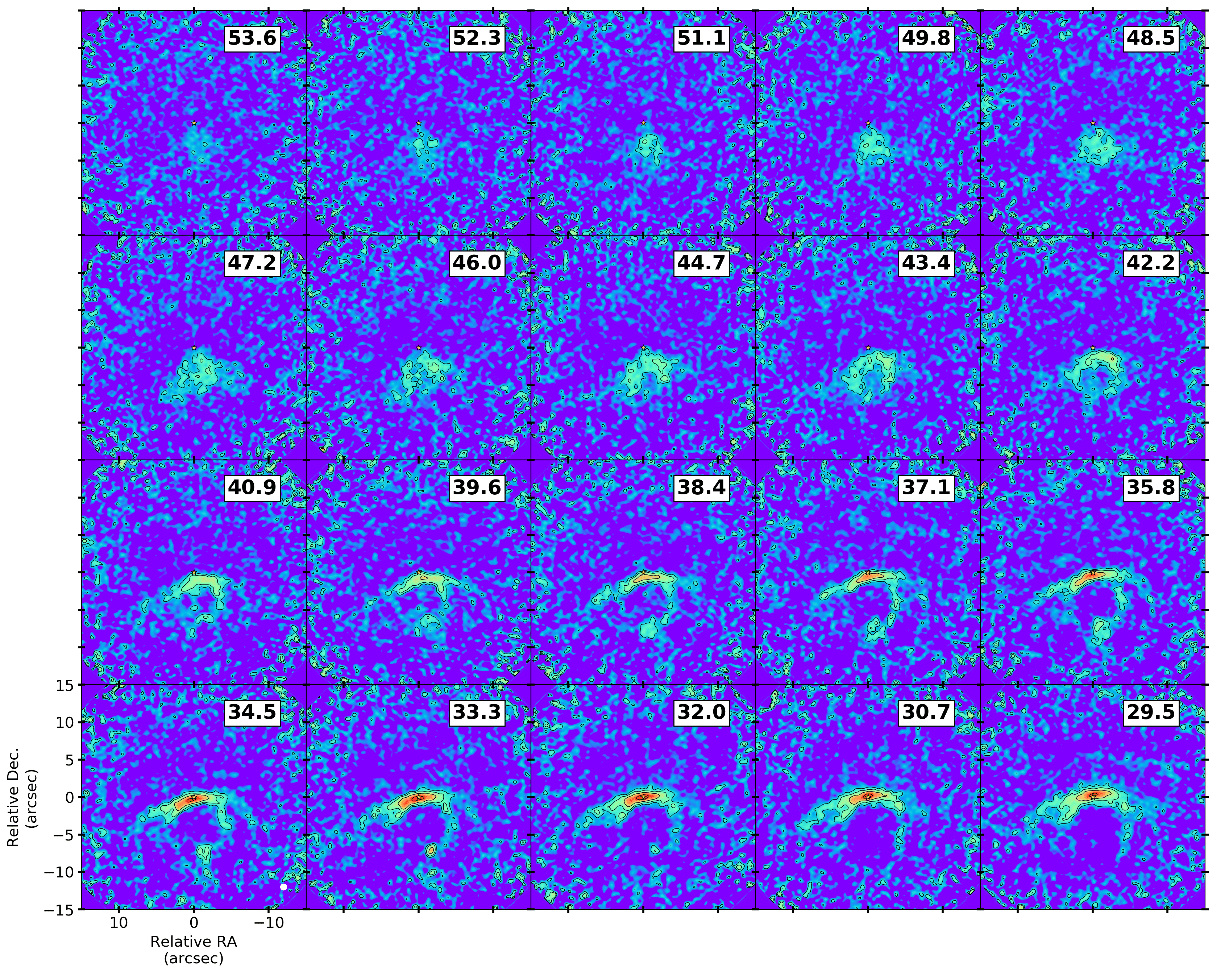}
		\includegraphics[width=14.cm]{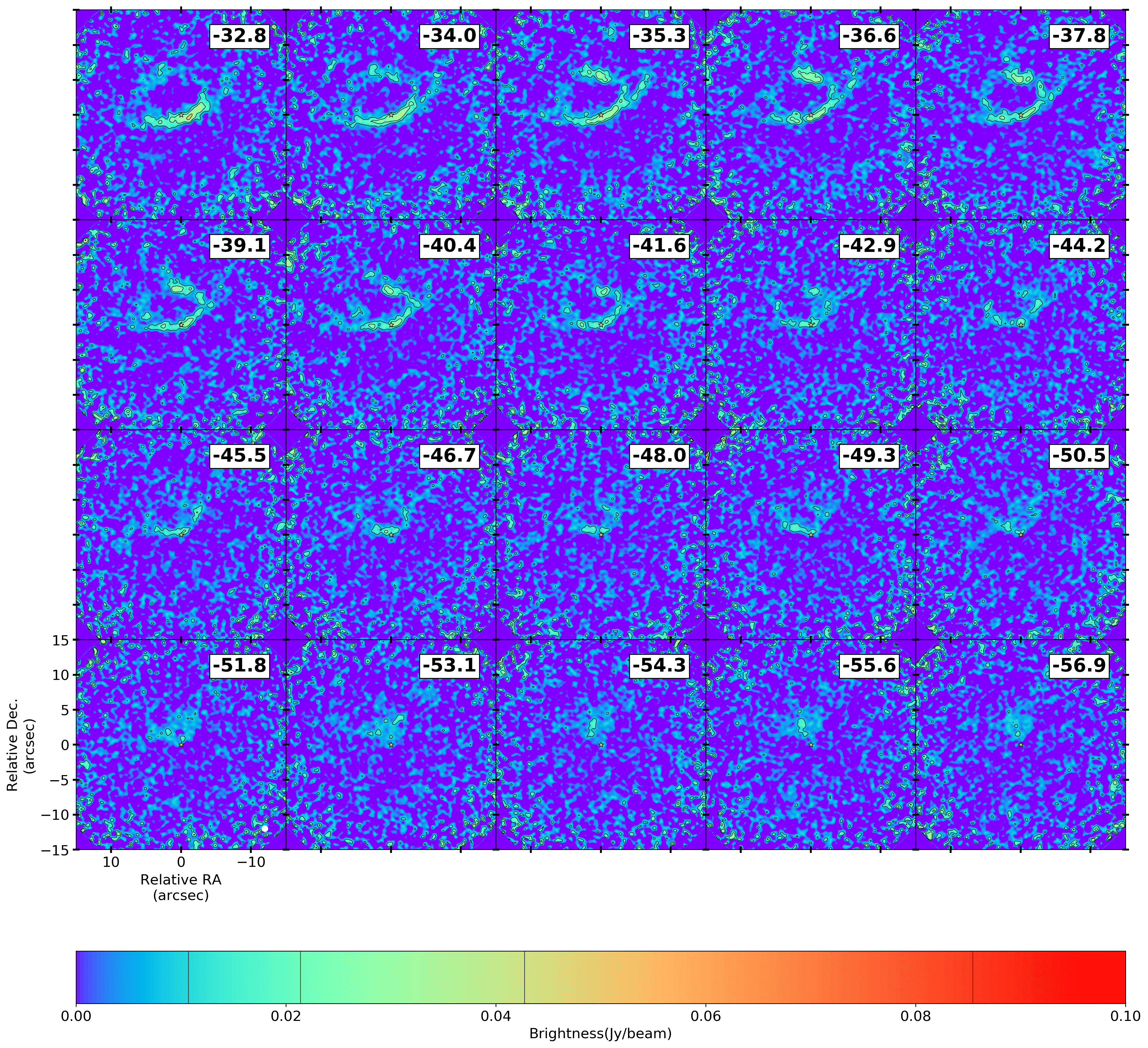}
		\caption{Channel maps of the red- (top) and blue- (bottom) shifted high-velocity wings of the CO emission. Image properties are identical to the left panel in Fig. \ref{COchan}.}
		\label{COwing}
	\end{figure*}

	Fig. \ref{COwing} shows the resolved emission distributions in the plane of the sky for the blue-shifted and red-shifted wings of the CO $J=\ $3$-$2 line, in the approximate $v_* \pm {\rm [30,55]} \kms$ velocity range. This is the velocity range for which \citet{Sahai1992} proposed the presence of high-velocity polar outflows. We detect signal above 3$\times \sigma_{\rm rms}$ up to maximal projected velocities of $-57\ \kms$ and $+54\ \kms$ with respect to the stellar velocity, speeds that are unusually high in comparison to typical AGB wind velocities of $\sim{\rm 5} - {\rm 20}\ \kms$ \citep[e.g.][]{Loup1993,Bujarrabal2001,Ramstedt2008,DeBeck2010,Danilovich2015}.
	Inspection of the emission patterns in the high-velocity tails reveals ring-like structures, as first detected by \citet{Doan2020}. Both line wings show emission features that start off as rather extended, bright, and only slightly curved arcs at lower speeds (around $v_* \pm$32~$\kms$). They have a length of $\sim$10'', and curve south(north)ward in the red- (blue-)shifted wing. These arcs progressively become more curved and compact as their speed increases. Notably at velocities around $v_* \pm {\rm 42}\ \kms$, the emission features become quasi-circular, with a distinct void in the center. The radius of these quasi-circles is $\sim$3'', and the width of the ring is $\sim$1'' measured from the 3$\times \sigma_{\rm rms}$ contour. These rings shrink and eventually become filled with emission as the highest speeds of the line wings are attained.
	
	Assuming only radial stream lines in the region where these features are detected, the observed patterns can be immediately recognised as thin shells of gas, each located at one respective pole, that are expanding into the polar direction at high speeds. The spectral evolution of the overall shape of the HVHG signal is similar in the blue and red-shifted line wings (see Fig. \ref{COwing}).	
	The signal in the red-shifted wing is notably stronger than its blue-shifted counterpart, by a factor $\sim$two. In addition, the signal in the (weaker) blue-shifted wing extends to speeds that are $\sim {\rm 3.5}\ \kms$ (7\%) larger than the opposing wing, though this may result from the data being limited to a surface brightness threshold. These bubbles are highly reminiscent of features that appear in more evolved post-AGB and planetary nebula systems, such as in e.g. Minkowski 92 \citep{Minkowski1946,Alcolea2007} or HD 101584 \citep{Humphreys1974,Olofsson2019}.
	
	\end{appendix}
	
\end{document}